\documentclass[preprint,3p,twocolumn,times,authoryear]{elsarticle}

\usepackage{pifont}
\usepackage{graphicx}
\usepackage{multirow}
\usepackage{array}
\usepackage{color}
\usepackage{setspace}
\usepackage{float} 
\usepackage{placeins}

\usepackage[table]{xcolor}
\usepackage{booktabs}
\usepackage{amssymb}
\usepackage{stfloats}
\usepackage{amsmath}
\usepackage{algorithm}
\usepackage{algpseudocode}
\usepackage{url}

\usepackage[colorlinks=true, citecolor=blue, linkcolor=blue, urlcolor=blue, bookmarksopen=false]{hyperref}
\begin{document}

\begin{frontmatter}

\title{Mutually Exclusive Multiclass Lesion Segmentation in Neuroimaging: Binary-Guided Weak Supervision with Inter-Class Orthogonality}

\author[1]{Ashutosh Kumar}
\ead{ashutosh3@iisc.ac.in}

\author[1]{Vivek Dhamale}
\ead{viveksdhamale@gmail.com}

\author[1]{Vaanathi Sundaresan\corref{cor1}}
\ead{vaanathi@iisc.ac.in}
\cortext[cor1]{Corresponding author}

\address[1]{Department of Computational and Data Sciences, Indian Institute of Science, Bengaluru 560012, Karnataka, India}

\begin{abstract}
Weakly supervised semantic segmentation (WSSS) offers a scalable alternative to voxel-level annotation in medical imaging, yet existing multiclass WSSS methods fail in clinically realistic settings where lesion subclasses co-occur spatially with ambiguous boundaries, producing overlapping activations and noisy pseudo-labels. This limitation is particularly pronounced in neuroimaging, where mutual exclusivity between pathological subregions is never explicitly enforced. We propose \textsf{BiMEx-MS} (Binary-guided Mutually Exclusive Multiclass Segmentation), a binary-guided framework that decomposes multiclass segmentation into whole-lesion localization and exclusive class assignment: a binary localization module provides a class-frequency-agnostic structural prior confining multiclass predictions within the detected lesion domain, while a multi-exit classification architecture with supervised contrastive pretraining produces multi-scale class-discriminative activation maps, which are aggregated via a class-specific attention network into mutually exclusive localization maps. Inter-class exclusivity is enforced through a tri-partite loss comprising per-class separation, inter-class orthogonality, and binary--multiclass spatial consensus, followed by hierarchical morphological pseudo-label refinement. Evaluated across brain tumor MRI (BraTS 2020, BraTS 2023 SSA) and intracranial hemorrhage CT (RSNA-ICH$\,\to\,$BHSD), \textsf{BiMEx-MS} significantly outperforms various weakly supervised baselines and achieves Edema HD95 of 29.56 mm on BraTS 2020 (the only method below 40 mm) and subdural hemorrhage Dice of 0.704 on BHSD, with gains consistently largest on boundary metrics and rare subtypes, confirming robustness to long-tailed class distributions. Extensive validation including cross-dataset and cross-continental generalization, detailed ablations, comparison across six backbone architectures, and uncertainty quantification confirm that structural guidance rather than model capacity drives performance, with calibrated uncertainty estimates supporting potential clinical deployment. Code is publicly available at \url{https://github.com/ashutoshkr45/BiMEx-MS-Neuro}.
\end{abstract}


\end{frontmatter}

\section{Introduction}
\label{sec:intro}

Medical image segmentation plays a central role in clinical diagnosis, treatment planning, and therapeutic decision-making. While recent development in deep learning has driven significant progress~\citep{zhou2019unetpp,yang2022efficient,hua2022symmetry}, supervised approaches require large volumes of annotated data, an important practical constraint in medical imaging where annotation is time-consuming and demands considerable clinical expertise. Weakly Supervised Semantic Segmentation (WSSS) has emerged as a prominent research direction to address this, leveraging lightweight supervision such as bounding boxes~\citep{dai2015boxsup,song2019box,lee2021bbam,wang2021bounding}, point annotations~\citep{bearman2016s,qu2019weakly,chen2021seminar}, and scribbles~\citep{gao2022segmentation,wu2023sparsely, lin2026adversarial}. Of various weak labels, image-level classification labels are lightweight and the most accessible, making them particularly appealing despite the additional challenges they introduce for model training~\citep{chikontwe2022weakly,pati2023weakly,chen2023ame}.

Most WSSS research addresses binary segmentation, separating a single target class from background, and the extension to multiclass segmentation is challenging and remains largely underexplored~\citep{chen2023ws,kuang2024weakly}. Multiclass medical segmentation is fundamentally harder: for instance, in glioma, rather than simply detecting tumor versus background, the task requires discriminating biologically distinct subregions (e.g., tumor core and edema) that exhibit both \textit{label symbiosis}, where they tend to co-occur, and \textit{spatial adjacency}, where they are typically contiguous or directly neighboring~\citep{patel2022weakly,kuang2024weakly}. These structural challenges are further compounded by severe class imbalance between dominant and rare subregions, long-tailed distributions (both in terms of frequency and size), and poor inter-class contrast that renders boundaries imperceptible on images. Hence, there is a need for tailored solutions beyond conventional binary approaches.

Existing WSSS methods fail characteristically in this setting. CAMs~\citep{zhou2016learning} are intrinsically biased toward discriminative regions; variants such as GradCAM~\citep{selvaraju2017grad} and LayerCAM~\citep{jiang2021layercam} improve binary localization but fail to resolve inter-class boundaries under class imbalance. Consistency-based methods~\citep{luo2020semi,gao2022segmentation,yang2025unimatch} expand activated regions via perturbation regularization but suffer the same per-class independence. Anatomically-constrained approaches~\citep{li2022deep} restrict the segmentation search space but do not generalize to pathological structures with high inter-patient variability. Transformer and foundation model-based methods~\citep{xu2022multi,xu2024mctformer+,lin2025semantic,yang2024foundation,zhu2026weaktr} achieve strong localization on natural images but rely on implicit attention-based separation that breaks down when lesion subregions share similar appearance characteristics. Even methods specifically targeting multiclass medical segmentation~\citep{chen2023ws,kuang2024weakly} rely on soft regularization that discourages but never strictly enforces pixel-wise inter-class exclusivity, without a unified whole-lesion boundary constraint. The common thread across all these directions is the absence of explicit spatial mutual exclusivity between foreground classes, producing overlapping activation maps for adjacent subregions (e.g., tumor core and edema).

To address these limitations, we propose \textsf{BiMEx-MS} (Binary-guided Mutually Exclusive Multiclass Segmentation), a weakly supervised multiclass segmentation framework that explicitly decouples whole-lesion localization from fine-grained inter-class discrimination. Our prior work \citep{dhamale2025inter} introduced the binary-guided mutual exclusivity principle as a proof of concept with minimal architectural formalization; \textsf{BiMEx-MS} provides a substantially extended and formally grounded treatment, mathematical formulation, and geometric interpretation of framework components including binary-guided refinement, aggregation network, and tri-partite loss functions introduced in~\cite{dhamale2025inter}. Here, as a new methodological contribution, we introduce hierarchical morphological pseudo-label refinement, and full segmentation network training on refined pseudo-labels, along with substantially expanded experimental validation. The main contributions of this extension are as follows:
\begin{itemize}
\item \textbf{Formal mathematical grounding of refinement:} Full breakdown of the binary-guided refinement with geometric interpretation, and explicit cosine-similarity log-forms with convergence arguments and expansion of contrastive loss in the multi-exit classification with contrastive pretraining, both absent in~\cite{dhamale2025inter}.

\item \textbf{Class-specific aggregation network:} Substantial expansion of aggregation network with class-conditioned importance weighting via class-specific input features, decomposed into explicit equations with formal normalization constraints. 

\item \textbf{Hierarchical pseudo-label refinement and segmentation:} Introduction of a morphological refinement strategy enforcing strict pixel-wise mutual exclusivity among foreground classes prior to full segmentation network training on refined pseudo-labels.

\item \textbf{Comprehensive multimodal and cross-dataset validation:} Addition of a new task on a different additional modality, leading to evaluation across two modalities and four datasets, against sixteen weakly supervised state-of-the-art baselines, including cross-dataset CT generalization and cross-continental MRI generalization and comparison across six backbone architectures.

\item \textbf{Robustness, scalability, and uncertainty analysis:} Extensive ablation studies isolating each component's contribution, progressive four-class expansion demonstrating graceful scalability with competitive tail class performance in long-tailed cases, and epistemic and aleatoric uncertainty quantification confirming anatomically meaningful confidence estimates, collectively establishing clinical reliability beyond binary overlap-based metrics.
\end{itemize}

\section{Related Work}
\label{sec:related}
Weakly supervised segmentation with image-level annotations has evolved from CAM-based localization to transformer and foundation model pipelines. Still, generating accurate dense predictions from global supervision remains particularly challenging in medical imaging due to ambiguous lesion boundaries, strong spatial co-occurrence between subregions, and absence of pixel-level annotations during training. This section reviews prior work in four directions: image-level weak supervision, CAM-based localization, transformer and foundation model-based methods, and finally, multiclass weakly supervised medical segmentation. 

\subsection{Weak Supervision Formulations Based on Image-Level Labeling}
\label{ssec:ws_imglabel}
Most WSSS methods are grounded in the Multiple Instance Learning (MIL) paradigm, where an image is supervised only at the image level. Early approaches rely on pixel affinity propagation and region growing~\citep{ahn2018learning,ahn2019weakly}, while global constraint-based methods~\citep{chen2022class,chen2023extracting} enforce image-level consistency to stabilize pixel-level predictions. Inter-image contrastive methods~\citep{xie2022c2am} improve foreground--background separation across samples. Recent advances include token-level pseudo-label co-training~\citep{yang2024weakly}, foundation model-enabled seed selection~\citep{yang2024foundation}, cross-modal vision-language prompting~\citep{lin2025semantic}, and uncertainty-aware pseudo-label refinement~\citep{zhang2025edge,fan2024pathmamba}; a comprehensive survey is provided by~\citet{chen2024weaklysupervisedsemanticsegmentationimagelevel}. While these methods advance image-level weakly supervised segmentation, they predominantly assume appearance-based inter-class contrast that is less predominant in neuroimaging, where co-occurring lesion subtypes share overlapping intensity distributions.

A key limitation is the implicit assumption of balanced and spatially separable classes, which is particularly problematic in medical imaging where rare substructures are significantly underrepresented. Loss reweighting strategies including focal loss~\citep{lin2017focal}, class-balanced loss~\citep{cui2019class,kervadec2019constrained}, and Dice- and Tversky-based objectives~\citep{milletari2016v,salehi2017tversky} address imbalance under full or partial supervision but do not resolve it under image-level weak supervision. More recent long-tailed recognition methods based on adaptive rebalancing~\citep{liu2019large}, decoupled representations~\citep{kang2019decoupling}, or logit adjustment~\citep{menon2020long} similarly assume pixel-level supervision, limiting applicability in pure WSSS settings. Consequently, MIL-based WSSS methods remain vulnerable to class imbalance, leading to poor representation of rare lesions.

\subsection{CAM-Based Localization and Pseudo-Label Refinement}
\label{ssec:cam_pseudolabel}
Class Activation Maps~\citep{zhou2016learning, chen2024weaklysupervisedsemanticsegmentationimagelevel} and their variants including GradCAM~\citep{selvaraju2017grad}, LayerCAM~\citep{jiang2021layercam}, ScoreCAM~\citep{wang2020score}, EigenCAM~\citep{muhammad2020eigen} remain the most widely used mechanism for weak localization. However, they suffer from partial activation, spatial bias toward discriminative regions, and coarse resolution due to deep downsampling. ReCAM~\citep{chen2022class} introduces class competition via reweighted activations; AME-CAM~\citep{chen2023ame} and ESFAN~\citep{zhang2025edge} improve resolution through multi-layer aggregation and edge-aware refinement; C2AM~\citep{xie2022c2am} improves feature separability via contrastive objectives.

A closely related direction involves pseudo-label refinement via teacher--student frameworks~\citep{liu2022perturbed,yang2025unimatch} and self-training~\citep{chen2021semi}. They iteratively refine noisy CAMs using confidence-based filtering, consistency regularization or moving average models. However, these strategies still remain sensitive to class imbalance, reinforcing long-tailed bias and suppressing rare lesion substructures. CAM refinement and loss reweighting thus improve localization quality but do not address the coupled challenges of class imbalance and structural overlap in multiclass weak supervision.

\subsection{Transformer and Foundation Model-Based Weak Supervision}
\label{ssec:trans_foundmodel}
Transformer and foundation model-based methods have extended weak supervision through attention-driven localization and large-scale visual-semantic pretraining. ViT-based WSSS methods introduce attention-driven localization. MCTformer~\citep{xu2022multi} leverages class tokens with patch affinity refinement, while MCTformer+~\citep{xu2024mctformer+} adds contrastive class token modules to ensure non-overlapping localization. 
WeakTr~\citep{zhu2026weaktr} improves class separation via adaptive attention fusion, while Sparse-ViT~\citep{hanna2023sparse} enforces attention head sparsity to retain semantically meaningful cues. CoSA~\citep{yang2024weakly} uses contrastive co-training with swapping assignments. Foundation model approaches extend weak supervision through large-scale pretraining: SemPLeS~\citep{lin2025semantic} enables CLIP-based semantic grounding, while SAM-based pipelines, FMA-WSSS~\citep{yang2024foundation} exploit class-agnostic segmentation priors for mask refinement. Recent medical adaptations~\citep{li2025review, ma2024segment} further demonstrate the potential of foundation models in low-label regimes. However, these methods improve representation richness without explicitly addressing structural class imbalance or enforcing constraints under multiclass lesion co-occurrence, leaving long-tailed effects unresolved.

\subsection{Structured and Multiclass Weakly Supervised Medical Segmentation}
\label{ssec:multiclass_seg}
Medical WSSS presents a fundamentally harder setting due to multiclass co-occurrence, hierarchical lesion structure, and severe class imbalance. WS-MTST~\citep{chen2023ws} extends transformer-based architectures to multi-label segmentation under implicit label continuity assumptions that are questionable for multifocal diseases (e.g., high-grade gliomas and intracranial hemorrhages). Feature decomposition approaches~\citep{kuang2024weakly} separate shared and class-specific representations via soft constraints but never strictly enforce pixel-wise inter-class exclusivity without a unified whole-lesion boundary constraint. Domain-adapted methods such as HAMIL~\citep{zhong2023hamil} and PathMamba~\citep{fan2024pathmamba} address resolution and long-range dependencies in histopathology but do not transfer to neuroimaging, where lesion subregions share similar intensity distributions and lack staining-based contrast. Loss engineering does not translate effectively to multiclass WSSS due to label symbiosis and the absence of reliable pixel-level supervision~\citep{kervadec2019constrained,patel2022weakly}, causing image-level imbalance to propagate into unstable pseudo-label generation.

Overall, existing methods are limited by the absence of explicit structural priors linking multiclass predictions to a shared lesion boundary, inability to enforce inter-class exclusivity at the pixel level, and failure to resolve long-tailed class imbalance without pixel-level supervision, producing discontinuous activations, overlapping predictions, and poor localization of rare lesion subregions.

Our prior work~\citep{dhamale2025inter} demonstrated that binary CAM can guide multiclass CAM learning via inter-class separability and agreement losses. \textsf{BiMEx-MS} substantially extends this by decoupling whole-lesion localization from inter-class discrimination and expanding a binary-guided structural prior that constrains multiclass predictions within a coherent lesion boundary, stabilizes learning under long-tailed distributions through a class-frequency-agnostic binary prior, and enforces inter-class orthogonality to prevent dominant classes from absorbing the activation budget of rare subtypes.

\section{Methodology}
\label{sec:method}
Weakly supervised segmentation from image-level labels is ill-posed, lacking explicit spatial or inter-class boundary constraints. This limitation is compounded in medical imaging by co-occurring lesion subregions with overlapping intensity distributions. \textsf{BiMEx-MS} addresses this by decomposing multiclass WSSS into coarse whole-lesion localization and fine-grained inter-class segmentation. We hypothesize that reliable binary lesion localization provides a structural prior that regularizes multiclass prediction, reduces inter-class ambiguity and enforces anatomically plausible segmentation under weak supervision.

\begin{figure*}[!t]
\centering
\includegraphics[width=1.00\textwidth,keepaspectratio]{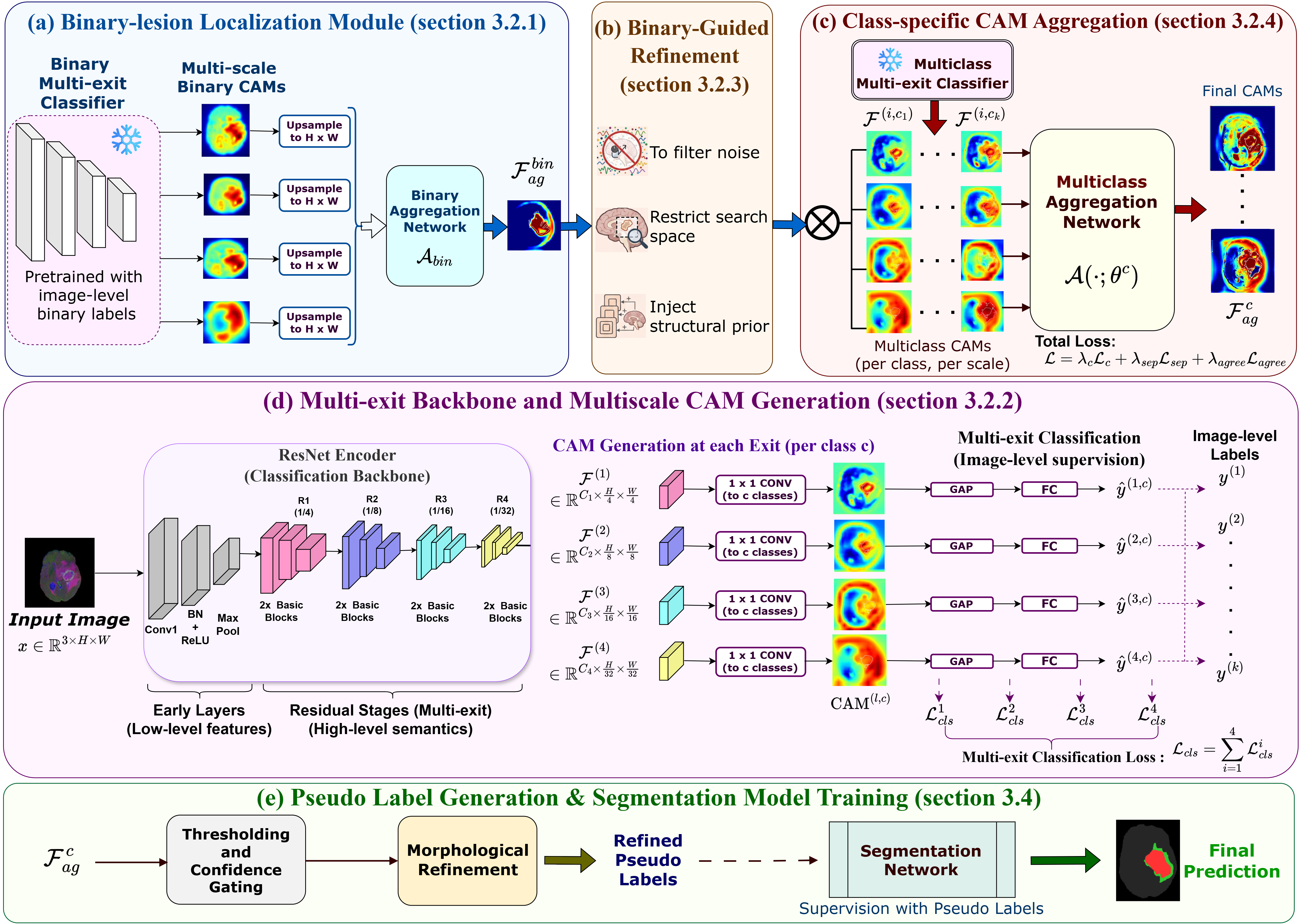}
    \caption{The proposed \textsf{BiMEx-MS} framework, \textbf{(a)} The binary lesion localization module produces the class-frequency-agnostic structural prior $\mathcal{F}_{\text{ag}}^{\text{bin}}$ via a frozen binary multi-exit classifier and binary aggregation network, \textbf{(b)} Binary-guided refinement, where the structural prior is injected at each scale via element-wise multiplication, confining multiclass competition to the lesion domain $\Omega^+$, \textbf{(c)} The class-specific aggregation network learns per-class spatial importance weights to produce mutually exclusive maps $\mathcal{F}_{\text{ag}}^{c}$,  \textbf{(d)} The multi-exit classification backbone (chose ResNet architecture as described in Section~\ref{ssec:arch_results}), instantiated separately for binary and multiclass streams, generates multiscale activation maps $\mathcal{F}^{(l,c)}$ at four scales under image-level supervision using $\mathcal{L}_{\text{cls}}$ (Eq.~\eqref{eq:cls}), \textbf{(e)} From each multiclass CAMs, confidence-gated predictions are obtained and their morphological refinement produces pseudo-labels supervising the final segmentation network.}
\label{fig:method}
\end{figure*}

\subsection{Problem Formulation}
\label{sec:formulation}

Let $\mathcal{X} \subset \mathbb{R}^{M \times H \times W}$ denote the input image space, with $M$ modalities of dimensions height ($H$) $\times$ width ($W$), and $\mathcal{Y} = \{0,1,\dots,C\}$ the label space, where $0$ denotes background and $c \in \{1,\ldots,C\}$ indexes foreground lesion subtypes. Let $\Omega = \{1,\ldots,H\} \times \{1,\ldots,W\}$ denote the pixel domain. A binary label is obtained by taking the logical union over all foreground classes:
\begin{equation}
    y_i^{\text{bin}} = \mathbf{1}\!\left[\,\exists\, c \in \{1,\ldots,C\}
    : y_i^c = 1\right].
    \label{eq:binlabel}
\end{equation}
Multiple classes co-occur within a single image ($\|\mathbf{y}_i\|_1 \geq 2$), introducing label symbiosis and spatial proximity that require mutual exclusivity in prediction. The training set is $\mathcal{D} = \{(x_i,\, y_i^{\text{bin}},\, \mathbf{y}_i)\}_{i=1}^{N}$, where $\mathbf{y}_i = [y_i^1, \ldots, y_i^C]^\top \in \{0,1\}^C$ is the multi-hot encoding vector for foreground classes, with no pixel-level annotation available at any stage.

The objective is to learn a segmentation function:
\begin{equation}
\Phi: \mathcal{X} \rightarrow \{0, 1, \ldots, C\}^{H \times W}
\end{equation}
that produces pixel-wise multiclass predictions $\hat{S}_i = \Phi(x_i)$ from image-level labels alone, with the mutual exclusivity constraint $\hat{S}_i(p) \in \{0, 1, \ldots, C\}$ enforcing that each pixel belongs to at most one foreground class.

\subsection{Overview of the Proposed Framework}
\label{sec:overview}
\textsf{BiMEx-MS} enforces mutual exclusivity via a binary-guided dual-module architecture, illustrated in Fig.~\ref{fig:method}: (i) a binary localization module that predicts the whole-lesion region as a structural prior, and (ii) a multiclass segmentation module that detects lesion subtypes via CAM-based localization constrained within this prior, followed by (iii) morphological pseudo-label refinement.

The binary localization problem is treated as foundational: $y_i^{\text{bin}}$ merges all foreground classes into a single positive category, producing a binary lesion mask $\hat{S}_i^{\text{bin}} \in \{0,1\}^{H \times W}$ that defines the restricted pixel domain $\Omega^+ = \{p \in \Omega : \hat{S}_i^{\text{bin}}(p) = 1\} \subsetneq \Omega$, within which multiclass competition is confined.

\subsubsection{Binary lesion localization module}
\label{sec:binary}
We train binary multi-exit classifier $f_{\text{bin}}:\mathcal{X}\rightarrow[0,1]$ to localize the whole-lesion extent using image-level label $y_i^{\text{bin}}$. It shares the same backbone and multi-exit architecture of the multiclass module (Section~\ref{ssec:multiclass}) but is parameterized for a single foreground class (binary classification) using image-level label $y_i^{\text{bin}}$. This decoupling is intentional: unlike the multiclass module, which must distinguish among lesion subtypes, $f_{\text{bin}}$ learns only the boundary between background and lesion, producing multi-scale activation maps $\{\mathcal{F}^{(l,\text{bin})}(x)\}_{l=1}^{4}$ (where $l \in \{1,2,3,4\}$ indexes the residual block exit) that are more spatially complete than class-specific activations $\mathcal{F}^{(l,c)}(x)$ (described in Section~\ref{ssec:multiclass}).

A binary aggregation network $\mathcal{A}_{\text{bin}}$, structurally identical to the multiclass aggregation module (Section~\ref{sec:aggregation}), combines these multi-scale activations into a single high-resolution prior:
\begin{equation}
\mathcal{F}_{\text{ag}}^{\text{bin}} = \mathcal{N}\!\left(\mathcal{A}_{\text{bin}}\!\left(
x, \{\mathcal{F}^{(l,\text{bin})}(x)\}_{l=1}^{4}\right)\right)\in[0,1]^{H\times W},
\label{eq:binprior}
\end{equation}
where $\mathcal{N}(\cdot)$ denotes min-max normalization. As shown in Fig.~\ref{fig:method}(a), $\mathcal{F}_{\text{ag}}^{\text{bin}}$ achieves broader lesion coverage than any individual $\mathcal{F}^{(l,c)}$, validating the decoupling strategy. At inference, $\Omega^+$ is enforced by thresholding $\mathcal{F}_{\text{ag}}^{\text{bin}}$, confining multiclass competition to the detected lesion region. Crucially, because $y_i^{\text{bin}}$ represents the union of all foreground classes irrespective of frequency, the resulting prior is class-frequency-agnostic, ensuring that both dominant and rare subtypes are equally represented within the shared lesion region.

\subsubsection{Multi-Scale class-discriminative activation maps using multi-exit architecture}
\label{ssec:multiclass}
CAM-based localization from a single convolutional layer is constrained by spatial resolution and insufficient for distinguishing clinically adjacent lesion subregions. Therefore, we adopt a multi-exit classification architecture~\citep{dhamale2025inter} with a $1{\times}1$ convolutional classifier after each of the four residual blocks $R_l$, $l \in \{1,2,3,4\}$, operating at resolutions $H_l \times W_l \in \left\{\frac{H}{4}{\times}\frac{W}{4},\, \frac{H}{8}{\times}\frac{W}{8},\, \frac{H}{16}{\times}\frac{W}{16},\, \frac{H}{32}{\times}\frac{W}{32}\right\}$ (Fig.~\ref{fig:method}(d)). Shallow exits capture fine local details, while deeper exits capture broader semantic context. Formally, let $\mathcal{F}^{(l,c)}(x) \in \mathbb{R}^{H_l 
\times W_l}$ denote the activation map for class $c$ at block $l$. At each exit, global average pooling (GAP) yields an image-level prediction:
\begin{equation}
    \hat{y}^{(l,c)} = \mathrm{GAP}\!\left(\mathcal{F}^{(l,c)}(x)\right)
    = \frac{1}{H_l W_l}\sum_{j=1}^{H_l}\sum_{k=1}^{W_l}
    \mathcal{F}^{(l,c)}_{jk}(x),
    \label{eq:gap}
\end{equation}
The network is trained jointly across all exits via a monotonically weighted loss:
\begin{equation}
    \mathcal{L}_{\text{cls}} = \sum_{l=1}^{4} \sum_{c}\, \beta_l \cdot
    \mathcal{L}_{\text{task}}\!\left(\hat{y}^{(l,c)},\, y^{(c)}\right),
    \label{eq:cls}
\end{equation}
where $\beta_l \in \{0.25, 0.50, 0.75, 1.0\}$ assign greater importance to deeper exits while preserving the spatial fidelity of shallower ones. For the binary module, $\mathcal{L}_{\text{task}}$ is binary cross-entropy and for the multiclass module, we use the multi-label focal loss:
\begin{equation}
    \mathcal{L}_{\text{focal}}^{(c)} = -\alpha_c \Bigl[
    y^{(c)}(1-p_c)^\gamma \log p_c 
    + (1-y^{(c)})\, p_c^\gamma \log(1-p_c)
    \Bigr],
    \label{eq:focal}
\end{equation}
where $p_c = sigmoid(\hat{y}^{(l,c)}) \in$ [0,1], $\alpha_c > 0$ is a per-class weight, and $\gamma \geq 0$ is the focusing parameter. For predicted probability $p_c$ and ground-truth label $y^{(c)}$, the focal weight $(1-p_c)^\gamma$ concentrates gradient mass on difficult examples, suppressing dominant classes and amplifying rare ones. This loss formulation produces a class-discriminative, multi-scale feature ensemble $\{\mathcal{F}^{(l,c)}(x)\}_{l=1}^{4}$ for subsequent aggregation.

\paragraph{Contrastive Pretraining.}
The multi-label objective in Eq.~\eqref{eq:cls} exerts no explicit pressure to separate co-occurring class representations, risking encoder collapse onto a shared subspace. To instill a discriminative inductive bias prior to multi-exit fine-tuning, both binary and multiclass encoders are pretrained via the multi-label supervised contrastive objective~\citep{khosla2020supervised}. Given two augmented views $\tilde{x}_i^{(1)}, \tilde{x}_i^{(2)}$ of image $x_i$ with projected representations $\mathbf{z}_i^{(1)}, \mathbf{z}_i^{(2)} \in \mathbb{R}^d$, the loss encourages embeddings sharing at least one class label to be attracted to each other:
\begin{equation}
    \mathcal{L}_{\text{con}} = -\frac{1}{|\mathcal{R}^+(i)|}
    \sum_{k \in \mathcal{R}^+(i)} \log
    \frac{\exp\!\left(\mathbf{z}_i \cdot \mathbf{z}_k \,/\, \rho\right)}
    {\displaystyle\sum_{a \in \mathcal{Q}(i)}
    \exp\!\left(\mathbf{z}_i \cdot \mathbf{z}_a \,/\, \rho\right)},
    \label{eq:contrastive}
\end{equation}
where $\mathcal{R}^+(i)$ is the set of positive pairs sharing at least one label with $x_i$, $\mathcal{Q}(i)$ is the full contrast set excluding $x_i$, and $\rho > 0$ is the temperature. The resulting representation geometry satisfies $\mathbf{z}_i \cdot \mathbf{z}_j \gg 0$ when $\mathbf{y}_i \odot \mathbf{y}_j \neq \mathbf{0}$ (i.e., at least one shared label) and $\mathbf{z}_i \cdot \mathbf{z}_j \ll 0$ otherwise, yielding class-specific features that are geometrically separated prior to fine-tuning.

\subsubsection{Binary-guided multiclass CAM refinement}
\label{sec:refinement}
Multiclass activation maps under image-level supervision suffer from two failure modes: \textit{background leakage}, where spurious activations occur at $p \in \Omega \setminus \Omega^+$; and \textit{inter-class blurring}, where class-$c$ activations appear at pixels belonging to $c' \neq c$. Both arise because Eq.~\eqref{eq:cls} imposes no spatial constraint on the support of $\mathcal{F}^{(l,c)}$ within $\Omega$. Here, we handle the above failure modes using $\mathcal{F}_{\text{ag}}^{\text{bin}}$, by integrating $\mathcal{F}_{\text{ag}}^{\text{bin}}$ and the multi-scale multiclass activation ensemble $\{\mathcal{F}^{(l,c)}(x)\}_{l=1}^{4}$. 

Prior to aggregation, each multiclass activation map is refined at its native scale by bilinear upsampling to $H \times W$ and element-wise multiplication with the normalized binary CAM, yielding the refined activation map $\tilde{\mathcal{F}}^{(l,c)}(x) \in \mathbb{R}^{H \times W}$:
\begin{equation}
    \tilde{\mathcal{F}}^{(l,c)}(x) = \mathcal{U}\!\left(\mathcal{F}^{(l,c)}(x)
    \right) \odot \mathcal{N}\!\left(\mathcal{F}_{\text{ag}}^{\text{bin}}\right),
    \label{eq:refinement}
\end{equation}
where $\odot$ denotes the Hadamard product, $\mathcal{N}(\cdot)$ min-max normalization to $[0,1]$, and $\mathcal{U}(\cdot)$ bilinear upsampling to $H \times W$. 
Eq.~\eqref{eq:refinement} acts as a spatially selective soft gate: $\tilde{\mathcal{F}}^{(l,c)}_{jk}(x) \approx 0$ for $p \in \Omega \setminus \Omega^+$ and $\tilde{\mathcal{F}}^{(l,c)}_{jk}(x) \approx \mathcal{F}^{(l,c)}_{jk}(x)$ for $p \in \Omega^+$, confining class-specific competition to the lesion domain $\Omega^+$ while preserving unconstrained intra-lesion discrimination.

This formulation contrasts with existing spatial compactness regularizers, which enforce contiguous single-region predictions per class, an assumption that fails in multifocal pathologies such as high-grade gliomas and intracranial hemorrhage. Eq.~\eqref{eq:refinement} imposes no topological constraint within $\Omega^+$, making the approach strictly more general for multi-focal neuroimaging settings.

\subsubsection{Class-Specific CAM aggregation}
\label{sec:aggregation}
Given the binary-guided ensemble $\{\tilde{\mathcal{F}}^{(l,c)}(x)\}_{l=1}^{4}$, the final step aggregates these maps into a single high-resolution map $\mathcal{F}_{\text{ag}}^{c} \in \mathbb{R}^{H \times W}$ per class. Uniform averaging is suboptimal: shallow layers $l \in \{1,2\}$ provide higher spatial resolution suited to compact classes, while deep layers $l \in \{3,4\}$ encode richer semantics suited to diffuse classes, and the ideal weighting is both class- and position-dependent.

We propose a \textit{class-specific attention-based aggregation network} $\mathcal{A}(\cdot;\theta^c)$, with learnable parameters $\theta^c$ that produces class-specific importance profiles driven by class-specific input features. During aggregation training, classifier backbones are frozen and only $\theta^c$ is updated. The input image $x$ is projected to a single-channel feature map via $P: \mathbb{R}^{M \times H \times W} \rightarrow \mathbb{R}^{H \times W}$ and masked by the normalized, refined activation map at each scale:
\begin{equation}
    \mathbf{m}^{(l,c)}(x) = P(x) \odot \mathcal{N}\!\left(\tilde{\mathcal{F}}^{(l,c)}(x)
    \right) \in \mathbb{R}^{H \times W}
    \label{eq:maskedinput}
\end{equation}
The masked inputs are concatenated channel-wise across all four exits and passed to $\mathcal{A}$ to produce pixel-wise importance scores:
\begin{equation}
    \tilde{w}^{(l,c)} = \mathcal{A}\!\left(
    \left[\mathbf{m}^{(l,c)}(x)\right]_{l=1}^{4};\,\theta^c\right),
    \label{eq:attention}
\end{equation}
where $[\cdot]_{l=1}^{4}$ denotes channel-wise concatenation over all blocks. The scores $\{\tilde{w}^{(l,c)}\}_{l=1}^{4}$ are normalized such that $\sum_{l=1}^{4} \tilde{w}^{(l,c)} = 1$ and $\tilde{w}^{(l,c)} \geq 0$ point-wise. The aggregated map is:
\begin{equation}
    \mathcal{F}_{\text{ag}}^{c}(x) = \sum_{l=1}^{4} \tilde{w}^{(l,c)} 
    \odot \tilde{\mathcal{F}}^{(l,c)}(x) \in \mathbb{R}^{H \times W}.
    \label{eq:aggmap}
\end{equation}

AME-CAM~\citep{chen2023ame} corresponds to the degenerate case, computing a single shared importance map across all classes. However, our network assigns a distinct importance profile $\{\tilde{w}^{(l,c)}\}_{l=1}^{4}$ per class, driven by class-specific input features, with profiles for $c \neq c'$ free to differ at every spatial location, which is particularly consequential for subtypes exhibiting maximal discriminability in multiclass setting at different spatial scales, providing a significant edge over AME-CAM.

\subsection{Loss Functions for Mutually Exclusive CAM Learning}
\label{sec:losses}
\begin{figure}
\centering
\includegraphics[width=1.00\columnwidth,keepaspectratio]{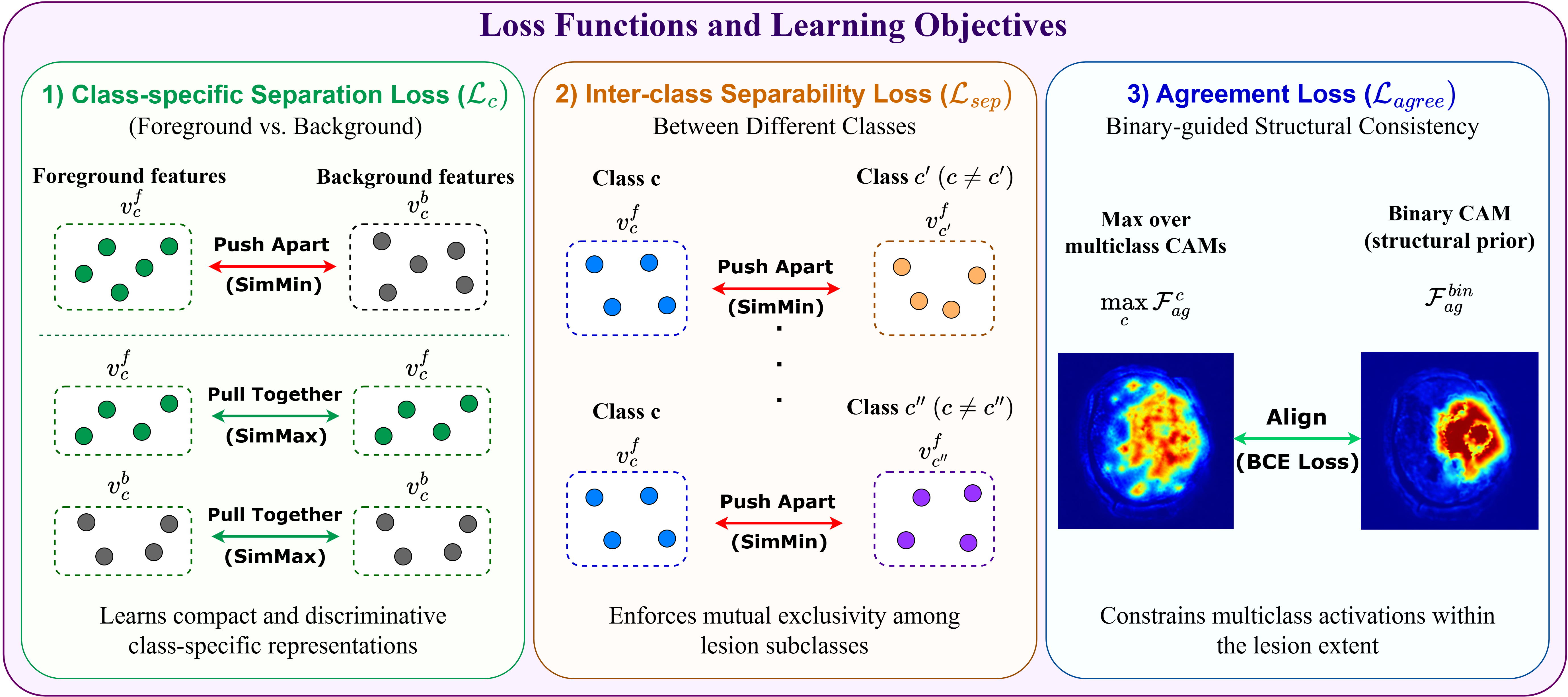}
\caption{Illustration of \textsf{BiMEx-MS} loss components for multiclass mutually exclusivity. The component  $\mathcal{L}_c$ (Eq.~\eqref{eq:lc}) enforces foreground--background separation, $\mathcal{L}_{\text{sep}}$ (Eq.~\eqref{eq:lsep}) ensures inter-class feature orthogonality and $\mathcal{L}_{\text{agree}}$ (Eq.~\eqref{eq:lagree}) provides binary-multiclass spatial consensus.}
\label{fig:loss}
\end{figure}

The aggregated maps $\mathcal{F}_{\text{ag}}^{c}$ are optimized under three complementary losses addressing per-class foreground--background separation, inter-class feature orthogonality, and spatial consistency with the binary prior, introduced in order of logical dependence (Fig.~\ref{fig:loss}).

\paragraph{Class-Specific Separation Loss $(\mathcal{L}_c)$.} This loss aims to provide good foreground--background discrimination. A $1{\times}1$ convolution $P: \mathbb{R}^{M \times H \times W} \rightarrow \mathbb{R}^{H \times W}$ projects to a single-channel feature map used to extract class-specific foreground and background representations:
\begin{equation}
    v_{f,c}^{s} = \mathcal{F}_{\text{ag}}^{c}(x^s) \odot P(x^s), \qquad
    v_{b,c}^{s} = \left(1 - \mathcal{F}_{\text{ag}}^{c}(x^s)\right) \odot P(x^s),
    \label{eq:features}
\end{equation}
where $v_{f,c}^{s}, v_{b,c}^{s} \in \mathbb{R}^{HW}$ are spatially flattened feature vectors for foreground and background, and $s$ indexes image instances. The separation loss is:
\begin{equation}
    \mathcal{L}_c = \sum_{c=1}^{C} \Bigl[
    \underbrace{\mathcal{L}_{\text{neg}}(v_{f,c}^{s},\, v_{b,c}^{t})}_{\text{push fg from bg}}
    + \underbrace{\mathcal{L}_{\text{pos}}(v_{f,c}^{s},\, v_{f,c}^{t})}_{\text{pull fg together}}
    + \underbrace{\mathcal{L}_{\text{pos}}(v_{b,c}^{s},\, v_{b,c}^{t})}_{\text{pull bg together}}
    \Bigr],
    \label{eq:lc}
\end{equation}
where $\mathcal{L}_{\text{neg}}(\mathbf{u},\mathbf{v}) = -\log(1 - \cos(\mathbf{u}, \mathbf{v}))$ minimizes cosine similarity and $\mathcal{L}_{\text{pos}}(\mathbf{u},\mathbf{v}) = -\log(\cos(\mathbf{u},\mathbf{v}))$ maximizes it, with similarities clamped to $[0,1]$. Minimization drives $\cos(v_{f,c}^{s}, v_{b,c}^{t}) \rightarrow 0$ while $\cos(v_{f,c}^{s}, v_{f,c}^{t}), \cos(v_{b,c}^{s}, v_{b,c}^{t}) \rightarrow 1$. $\mathcal{L}_c$ is applied per class rather than cumulatively to prevent a common foreground representation from being imposed across classes.

\paragraph{Inter-Class Separability Loss $(\mathcal{L}_{\text{sep}})$.}
Per-class separation alone does not guarantee mutual exclusivity: even when $\mathcal{L}_c = 0$ for all $c$, foreground representations of distinct classes may satisfy $\cos(v_{f,c'}^{s}, v_{f,c}^{t}) \approx 1$, activating both classes at the same pixels. This representational collapse is the fundamental failure mode of weakly supervised multiclass segmentation, wherein the model combines lesion detection with subtype attribution at the pixel level. To resolve this, we penalize pairwise cosine similarity between foreground representations of all class pairs:
\begin{align}
    \mathcal{L}_{\text{sep}} &= \sum_{\substack{c,c'=1\\c \neq c'}}^{C}
    \mathcal{L}_{\text{neg}}\!\left(v_{f,c'}^{s},\, v_{f,c}^{t}\right)\\
    &=-\sum_{\substack{c,c'=1\\c \neq c'}}^{C}
    \log\!\left(1 - \cos(v_{f,c'}^{s},\, v_{f,c}^{t})\right),  
    \label{eq:lsep}
\end{align}
The above minimization drives $\cos(v_{f,c'}^{s}, v_{f,c}^{t}) \rightarrow 0\ \forall\ c \neq c'$ and constrains spatial vectors to be mutually orthogonal, preventing any pixel from simultaneously contributing to two distinct foreground classes. This orthogonality also defends against class-imbalance-driven dominance: without $\mathcal{L}_{\text{sep}}$, a high-frequency class with large $\|v_{f,c}^s\|$ may align its representation with a rare class, absorbing its activation budget. Together, $\mathcal{L}_c$ separates each class from its background while $\mathcal{L}_{\text{sep}}$ partitions the foreground feature space into $C$ mutually exclusive subspaces.

\paragraph{Agreement Loss $(\mathcal{L}_{\text{agree}})$.}
With separability enforced, the union of multiclass predictions must align with the binary prior to prevent under-coverage of $\Omega^+$ or leakage into $\Omega \setminus 
\Omega^+$. We impose this alignment as follows:
\begin{equation}
    \mathcal{L}_{\text{agree}} = \mathcal{L}_{\text{BCE}}\!\left(
    \max_{c \in \{1,\ldots,C\}}\, \mathcal{F}_{\text{ag}}^{c}(x);\;
    \operatorname{sg}\!\left(\mathcal{F}_{\text{ag}}^{\text{bin}}(x)\right)\right),
    \label{eq:lagree}
\end{equation}
where $\operatorname{sg}(\cdot)$ denotes stop-gradient preventing back-propagation through binary module. Geometrically, $\mathcal{L}_{\text{agree}}$ minimizes the symmetric difference between the predicted multiclass union and the binary lesion map, penalizing both under-coverage ($\max_c \mathcal{F}_{\text{ag}}^{c}(p) \approx 0$ where $\mathcal{F}_{\text{ag}}^{\text{bin}}(p) \approx 1$) and leakage ($\max_c \mathcal{F}_{\text{ag}}^{c}(p) \approx 1$ where $\mathcal{F}_{\text{ag}}^{\text{bin}}(p) \approx 0$).  Crucially, while $\mathcal{L}_{\text{sep}}$ enforces orthogonal feature directions but imposes no lower bound on activation magnitude within $\Omega^+$, $\mathcal{L}_{\text{agree}}$ ensures that at least one class activates wherever the binary prior detects a lesion, irrespective of training frequency. Together with the element-wise refinement of Eq.~\eqref{eq:refinement}, this loss closes the feedback loop between modules, coupling the structural prior to the multiclass output (via $\mathcal{L}_c$, $\mathcal{L}_{\text{sep}}$).

\paragraph{Total Loss.}
The three components are combined as,
\begin{equation}
    \mathcal{L} = \lambda_c\, \mathcal{L}_c + \lambda_{\text{sep}}\,
    \mathcal{L}_{\text{sep}} + \lambda_{\text{agree}}\, \mathcal{L}_{\text{agree}},
    \label{eq:total}
\end{equation}
where $\lambda_c, \lambda_{\text{sep}}, \lambda_{\text{agree}} > 0$ are scalar weights. The three terms perform distinct roles: $\mathcal{L}_c$ enforces per-class foreground--background separability, $\mathcal{L}_{\text{sep}}$ enforces inter-class orthogonality, and $\mathcal{L}_{\text{agree}}$ enforces global spatial consistency with the binary prior. The necessity of each is further examined in Section~\ref{sssec:loss_abl}.

\subsection{Pseudo-Label generation and final segmentation}
\label{sec:postprocess}
After training, activation maps are converted into segmentation predictions through three sequential steps: confidence-gated thresholding, morphological pseudo-label refinement and pseudo-label-supervised segmentation.

\paragraph{Thresholding and Class-Presence Suppression.}
At inference, $\mathcal{F}_{\text{ag}}^{\text{bin}}$ refines each multiclass activation via Eq.~\eqref{eq:refinement} before aggregation into $\mathcal{F}_{\text{ag}}^{c}$. 
A global threshold $\tau_{\text{bin}}$ suppresses multiclass activations at background pixels (classified as background by $\mathcal{F}_{\text{ag}}^{\text{bin}}$):
\begin{equation}
    \tilde{\mathcal{F}}_{\text{ag}}^{c}(p) =
    \begin{cases}
        \mathcal{F}_{\text{ag}}^{c}(p) \cdot \mathcal{F}_{\text{ag}}^{\text{bin}}(p) 
        & \text{if } \mathcal{F}_{\text{ag}}^{\text{bin}}(p) > \tau_{\text{bin}}, \\
        0 & \text{otherwise,}
    \end{cases}
    \label{eq:binthresh}
\end{equation}
followed by a class-specific threshold $\tau_c$ yielding a class-wise initial hard prediction:
\begin{equation}
    \hat{S}^{c}(p) = \mathbf{1}\!\left[\tilde{\mathcal{F}}_{\text{ag}}^{c}(p) > \tau_c\right].
    \label{eq:classthresh}
\end{equation}
A confidence gate additionally suppresses all predictions for class $c$ if $sigmoid(\hat{y}^{(4,c)}) < \tau_{\text{conf}}$, preventing absent classes from generating spurious spatial activations under multi-label conditions.

\paragraph{Morphological refinement of pseudo-labels.}
Raw thresholded maps exhibit structural artifacts (intra-lesion holes, boundary discontinuities and isolated spurious activations) that degrade pseudo-label quality. Rather than refining each class mask independently, we adopt a hierarchical assignment strategy: the union of all predicted class masks is consolidated into a single lesion region and refined jointly via hole filling, morphological closing, and noise suppression. During morphological filling of holes, in disputed pixels, class assignment prioritizes the class with the smallest extent. This ensures strict mutual exclusivity while preventing independent refinement from distorting inter-class boundaries.

\begin{algorithm}[p]
\caption{Inference Pipeline of \textsf{BiMEx-MS}}
\label{alg:inference}
\begin{algorithmic}[1]
\footnotesize
\Require Test image $x \in \mathbb{R}^{M \times H \times W}$, frozen binary module, frozen multi-exit classifier, frozen aggregation networks $\{\mathcal{A}(\cdot;\theta^c)\}_{c=1}^{C}$, frozen segmentation model trained on pseudo-labels; thresholds 
$\tau_{\text{bin}},\, \{\tau_c\}_{c=1}^{C},\, \tau_{\text{conf}}$
\Ensure Segmentation map $\hat{S}_{\text{final}} \in 
\{0,\ldots,C\}^{H \times W}$

\Statex
\State \textit{// Stage 1: Binary prior}
\State $\mathcal{F}_{\text{ag}}^{\text{bin}}(x) \leftarrow 
\mathcal{N}\!\left(\textsc{BinaryStream}(x)\right)$

\Statex
\State \textit{// Stage 2: Multiclass map extraction}
\State Extract $\{\mathcal{F}^{(l,c)}(x)\}_{l=1}^{4}$ from 
multi-exit classifier, $\forall\, c \in \{1,\ldots,C\}$

\Statex
\State \textit{// Stage 3: Per-class refinement, aggregation, 
and thresholding}
\For{$c = 1, \ldots, C$}
    \If{$sigmoid(\hat{y}^{(4,c)}) < \tau_{\text{conf}}$}
        \State $\hat{S}^{c}(p) \leftarrow 0 \quad \forall\, p$
        \hfill $\triangleright$ suppress absent class
    \Else
        \For{$l = 1, \ldots, 4$}
            \State $\tilde{\mathcal{F}}^{(l,c)}(x) \leftarrow
            \mathcal{U}\!\left(\mathcal{F}^{(l,c)}(x)\right) \odot
            \mathcal{N}\!\left(\mathcal{F}_{\text{ag}}^{\text{bin}}
            \right)$
            \hfill $\triangleright$ Eq.~\eqref{eq:refinement}
            \State $\mathbf{m}^{(l,c)}(x) \leftarrow P(x) \odot
            \mathcal{N}\!\left(\tilde{\mathcal{F}}^{(l,c)}(x)\right)$
            \hfill $\triangleright$ Eq.~\eqref{eq:maskedinput}
        \EndFor
        \State $\{\tilde{w}^{(l,c)}\}_{l=1}^{4} \leftarrow
        \mathcal{A}\!\left(\left[\mathbf{m}^{(l,c)}(x)
        \right]_{l=1}^{4};\,\theta^c\right)$,\;
        \hspace{2cm}normalised s.t.\ $\sum_{l=1}^{4}\tilde{w}^{(l,c)} = 1$
        \hfill $\triangleright$ Eq.~\eqref{eq:attention}
        \State $\mathcal{F}_{\text{ag}}^{c}(x) \leftarrow
        \sum_{l=1}^{4} \tilde{w}^{(l,c)} \odot
        \tilde{\mathcal{F}}^{(l,c)}(x)$
        \hfill $\triangleright$ Eq.~\eqref{eq:aggmap}
        \For{each pixel $p \in \Omega$}
            \If{$\mathcal{F}_{\text{ag}}^{\text{bin}}(p) >
            \tau_{\text{bin}}$}
                \State $\tilde{\mathcal{F}}_{\text{ag}}^{c}(p)
                \leftarrow \mathcal{F}_{\text{ag}}^{c}(p) \cdot
                \mathcal{F}_{\text{ag}}^{\text{bin}}(p)$
                \hfill $\triangleright$ Eq.~\eqref{eq:binthresh}
            \Else
                \State $\tilde{\mathcal{F}}_{\text{ag}}^{c}(p)
                \leftarrow 0$
            \EndIf
        \EndFor
        \State $\hat{S}^{c}(p) \leftarrow \mathbf{1}\!\left[
        \tilde{\mathcal{F}}_{\text{ag}}^{c}(p) > \tau_c
        \right] \quad \forall\, p$
        \hfill $\triangleright$ Eq.~\eqref{eq:classthresh}
    \EndIf
\EndFor

\Statex
\State \textit{// Stage 4: Morphological refinement and exclusive 
assignment}
\State $\Omega^{+}_{\text{pred}} \leftarrow
\textsc{MorphRefine}\!\left(\bigcup_{c=1}^{C}\hat{S}^{c}\right)$
\State Assign class masks hierarchically within 
$\Omega^{+}_{\text{pred}}$ in ascending order of spatial extent
\State $\hat{S}_{\text{final}} \leftarrow 
\textsc{ExclusiveAssign}\!\left(\{\hat{S}^{c}\}_{c=1}^{C},\,
\Omega^{+}_{\text{pred}}\right)$

\Statex
\State \textit{// Stage 5: Segmentation model prediction}
\State \Return $\displaystyle\operatorname*{arg\,max}_{c \in \{0,\ldots,C\}}\;
\hat{O}^{c}_{p}\!\left(\textsc{SegModel}(x)\right), \quad \forall\, p$
\Statex \hspace{0.5cm} $\triangleright$ pixel-wise class with highest softmax probability

\end{algorithmic}
\end{algorithm}

\paragraph{Segmentation model training on pseudo-labels.}
The refined pseudo-labels provide per-pixel supervision for a Wide-ResNet-38 segmentation model~\citep{wu2019wider},  tailored for dense prediction tasks with widespread application in weakly supervised segmentation frameworks \citep{xu2024mctformer+} due to its ability to maintain a multiscale receptive field while avoiding resolution reduction. We train the model to predict $C{+}1$ classes (including background) with a frozen pretrained backbone and prediction heads trained from scratch. The training objective combines cross-entropy and Dice losses:
\begin{equation}
    \mathcal{L}_{\text{seg}} = \mathcal{L}_{\text{CE}}\!\left(\hat{O},\, 
    \hat{S}_{\text{final}}\right) + \mathcal{L}_{\text{Dice}}\!\left(\hat{O},\, 
    \hat{S}_{\text{final}}\right),
    \label{eq:segloss}
\end{equation}
where $\hat{O}$ is the softmax output, $\mathcal{L}_{\text{CE}}$ provides dense pixel-level supervision, and $\mathcal{L}_{\text{Dice}}$ mitigates class-frequency imbalance across lesion subtypes.
The complete inference pipeline is summarized in Algorithm~\ref{alg:inference}.

\section{Experimental Setup}

\subsection{Dataset Details}
\label{sec:datasets}

We evaluate \textsf{BiMEx-MS} across four datasets spanning two imaging modalities, two neurological pathologies, and two experimental protocols (in-domain evaluation and cross-dataset generalization). Table~\ref{tab:datasets} summarizes slice-level statistics for all datasets.
\begin{table}[t]
\centering
\caption{Summary of datasets used. Image-level labels are used for training; voxel-level annotations are used for evaluation only. Slices from the same subject are strictly confined to a single split to avoid data leakage across splits.}
\label{tab:datasets}
\scriptsize
\setlength{\tabcolsep}{4pt}
\renewcommand{\arraystretch}{1.0}
\begin{tabular}{llrrrrr}
\toprule
\multicolumn{7}{c}{\textbf{MRI -- Brain Tumor}} \\
\cmidrule(r){1-7}
\textbf{Dataset} & \textbf{Split}
& \textbf{Total} & \textbf{Core} & \textbf{Edema} & & \\
\midrule
\multirow{3}{*}{BraTS 2020}
& Train & 36,735 & 10,324 & 15,580 & & \\
& Val   &  9,145 &  2,751 &  3,846 & & \\
& Test  & 11,315 &  3,669 &  4,713 & & \\
\cmidrule{2-5}
& \textit{Total} & \textit{57,195} & \textit{16,744} & \textit{24,139} & & \\
\midrule
BraTS 2023 SSA & Test & 9,300 & 3,110 & 4,426 & & \\
\bottomrule
\toprule
\multicolumn{7}{c}{\textbf{CT -- Intracranial Haemorrhage}} \\
\cmidrule(r){1-7}
\textbf{Dataset} & \textbf{Split}
& \textbf{Total} & \textbf{EPH} & \textbf{IPH}
& \textbf{IVH} & \textbf{SDH} \\
\midrule
\multirow{2}{*}{RSNA-ICH}
& Train & 82,564 & 2,074 & 17,813 & 14,969 & 18,561 \\
& Val   & 20,642 &   502 &  4,403 &  3,664 &  4,579 \\
\cmidrule{2-7}
& \textit{Total} & \textit{103,206} & \textit{2,576}
  & \textit{22,216} & \textit{18,633} & \textit{23,140} \\
\midrule
\multirow{3}{*}{BHSD}
& Train &  4,086 &  123 &   643 &   524 &   568 \\
& Val   &    681 &   19 &    94 &    70 &    70 \\
& Test  &  1,209 &   39 &   151 &   119 &   127 \\
\cmidrule{2-7}
& \textit{Total} & \textit{5,976} & \textit{181}
  & \textit{888} & \textit{713} & \textit{765} \\
\bottomrule
\end{tabular}
\end{table}

\paragraph{Brain Tumor Segmentation (BraTS 2020).}
The BraTS 2020 dataset~\citep{menze2014multimodal,bakas2017advancing,bakas2018identifying} is publicly available and comprises multimodal preoperative MRI scans from 369 glioma patients; We use T1ce, T2, and FLAIR sequences as the three-channel input. Voxel-level annotations delineate three tumor subregions consolidated into two foreground classes (\textit{Tumor Core}, \textit{Edema}) following standard practice. Image-level labels are derived from voxel-wise labels and are used solely for training, and  voxel-level masks are reserved for evaluation. For an image-level label of a class to be positive, there must be at least one foreground voxel in the slice belonging to the respective class. The dataset is partitioned into 237/59/73 training/validation/test patients (correspondng slice-level splits are indicated in Table~\ref{tab:datasets}), ensuring strict subject-level isolation with no data leakage across splits. 

\paragraph{BraTS 2023 Sub-Saharan Africa (BraTS 2023 SSA).}
For brain tumor cross-data generalization experiment, the BraTS 2020-trained model is evaluated directly on the BraTS 2023 SSA dataset~\citep{adewole2023brain}, publicly available and comprising 60 glioma volumes acquired across Sub-Saharan African institutions using the sequences, protocol and labeling same as BraTS 2020, with no shared patients, institutions, or acquisition protocols. No fine-tuning is performed on the dataset.

\paragraph{RSNA Intracranial Hemorrhage Detection (RSNA-ICH).}
The RSNA-ICH dataset~\citep{rsna-intracranial-hemorrhage-detection} publicly available and contains over 25,000 non-contrast head CT studies with multi-label image-level annotations. Subarachnoid hemorrhage (SAH) is excluded due to its diffuse sulcal distribution, leaving four foreground classes: Intraparenchymal (IPH), Intraventricular (IVH), Subdural (SDH), and Epidural (EPH). Three-channel windowing (brain windowing: center/width - 30/80~HU; subdural windowing: center/width - 80/200~HU; bone windowing: center/width - 600/2800~HU), yielding a 3-channel input resized to $224{\times}224$ pixels. 
After stratified sampling, training and validation splits contain 82,564 and 20,642 images respectively. EPH is severely underrepresented (refer Table~\ref{tab:datasets}, $\sim$2.5\% of total slices), reflecting the long-tailed distribution in real-world hemorrhage datasets; targeted EPH augmentation (rotation $\pm10^\circ$, affine scaling and translation, brightness/contrast jitter, Gaussian noise $\sigma \in [0.01, 0.05]$, and gamma correction via Albumentations~\citep{buslaev2020albumentations}) is applied to partially mitigate this imbalance at the training stage. As no voxel-level annotations are available, RSNA-ICH is used exclusively for weakly supervised training.

\paragraph{Brain Hemorrhage Segmentation Dataset (BHSD).}
The BHSD dataset~\citep{wu2023bhsd} is publicly available consists of 192 CT volumes with voxel-level annotations, across the same four subtypes, yielding 5,976 axial slices after RSNA-consistent windowing and 3-channel decomposition. For within-data evaluation, volumes are partitioned into 132/21/39 splits with strict subject-level isolation to prevent data leakage across subjects (refer to Table~\ref{tab:datasets} for slice-level split). For ICH cross-data generalization, we train on RSNA-ICH image-level labels and evaluate using the BHSD test subset voxel-level masks.

\subsection{Implementation Details}

All models are implemented in PyTorch and trained on NVIDIA A6000 GPUs. Both encoders undergo contrastive pretraining~\citep{khosla2020supervised} prior to multi-exit fine-tuning. For BraTS, pretraining runs for 100 and 50 epochs (binary and multiclass, respectively); classifiers are subsequently fine-tuned for 50 epochs using Adam optimizer~\citep{kingma2014adam} ($lr = 5\times10^{-4}$ for multiclass, $1\times10^{-3}$ for binary, batch size $=$ 155), and aggregation networks for 50 epochs at $1\times10^{-3}$. For ICH, classifiers are trained for 80 epochs (determined using early stopping with patience of 10 validation cycles) using batch size of 128 and aggregation networks for 50--100 epochs at $5\times10^{-4}$. Loss weights are fixed empirically using trial-and-error method at $\lambda_c = 1$, $\lambda_{\text{sep}} = 1$, $\lambda_{\text{agree}} = 5$. The Wide-ResNet38 segmentation model is trained on pseudo-labels for 30 epochs at $2\times10^{-3}$ with polynomial LR decay. 

\subsection{Evaluation metrics and Statistical testing}
\label{ssec:eval_metrics}

Segmentation performance is quantified using Dice Similarity Coefficient (Dice), Average Symmetric Surface Distance (ASSD)~\citep{heimann2009comparison} in mm, and 95th Percentile Hausdorff Distance (HD95)~\citep{menze2014multimodal} in mm. Dice measures overlap agreement between predicted and ground-truth masks. ASSD computes the mean of all pairwise distances between predicted and ground-truth boundary surfaces, measured symmetrically in both directions:
\begin{equation}
\begin{split}
    \text{ASSD}(M_{\text{pred}}, G) = \frac{1}{|M_{\text{pred}}| + |G|}
    \Biggl(
    &\sum_{m \in \partial M_{\text{pred}}} \min_{g \in \partial G} \|m - g\|_2 \\
    + &\sum_{g \in \partial G} \min_{m \in \partial M_{\text{pred}}} \|g - m\|_2
    \Biggr),
\end{split}
    \label{eq:assd}
\end{equation}
where $\partial M_{\text{pred}}$ and $\partial G$ denote the boundary point sets of the predicted and ground-truth masks respectively, and $|\cdot|$ denotes cardinality. 
ASSD is particularly sensitive to systematic boundary biases arising from inter-class overlap (the central failure mode addressed by \textsf{BiMEx-MS}). HD95 computes the 95th percentile of the directed Hausdorff distance distribution, capturing near-worst-case boundary deviations while remaining robust to isolated outlier vertices. Together, these three metrics provide a clinically meaningful and non-redundant evaluation profile. Statistical comparisons between \textsf{BiMEx-MS} and the strongest baseline are performed using the Wilcoxon signed-rank test with identical seeds to ensure paired observations; differences are considered significant at $p < 0.05$.

\subsection{Experiments}

\subsubsection{Comparison of encoder backbone architectures}
\label{sssec:arch_comp}
We compare six encoder backbones: ResNet-18 and ResNet-50~\citep{he2016deep}, DenseNet-121~\citep{huang2017densely}, MobileNet-V2~\citep{sandler2018mobilenetv2}, EfficientNet-B1~\citep{tan2019efficientnet}, and Swin-Tiny~\citep{liu2021swin}, spanning 3.4M--28.9M parameters (Fig.~\ref{fig:backbone_analysis} for exact numbers) based on the metrics defined in Section~\ref{ssec:eval_metrics}. The best-performing backbone is selected for subsequent experiments. 

\subsubsection{Comparison with weakly supervised state-of-the-art methods}
\label{sssec:sota}
We evaluated \textsf{BiMEx-MS} against four groups of weakly supervised segmentation methods using the metrics described in Section~\ref{ssec:eval_metrics}. (i) \textit{Classical CAM-based methods}: GradCAM, LayerCAM, ScoreCAM, EigenCAM, and ReCAM; (ii) \textit{Medical WSSS methods}:  AME-CAM, HAMIL, ESFAN, and PathMamba; (iii) \textit{Transformer-based methods}: MCTformer, MCTformer+, Sparse-ViT, CoSA, and WeakTr; (iv) \textit{Foundation model-assisted methods}: SemPLeS and FMA-WSSS (using CLIP and SAM priors for semantic grounding and boundary refinement). Note that methods designed exclusively for single-class segmentation or requiring modality-specific foundation model adaptation are excluded from comparison, as their extension to the multiclass neuroimaging setting would introduce confounds beyond a controlled evaluation. Our framework performance is reported for both ResNet-18 and ResNet-50 backbones. Baseline hyperparameters are matched to our training protocol where applicable, with remaining hyperparameters adopted from their respective publications (or tuned to achieve better performance for a fair comparison).

\subsubsection{Cross-Dataset Generalization}
\label{sssec:crossdata_gen}
For brain tumor segmentation, the BraTS 2020-trained model is evaluated directly on BraTS 2023 SSA without any fine-tuning. This constitutes a strict cross-continental generalization test across independent patients, acquisition protocols, or imaging sites. For ICH, we train exclusively on RSNA-ICH image-level labels and evaluate against BHSD voxel-level masks, assessing generalization across independently acquired CT datasets.

\subsubsection{Evaluation under progressive class expansion}
\label{sssec:class_exp}
Scalability of \textsf{BiMEx-MS} is evaluated within the cross-dataset RSNA$\,\to\,$BHSD configuration by incrementally introducing hemorrhage subtypes: IPH\,+\,SDH $\to$ IPH\,+\,SDH \,+\,IVH $\to$ IPH\,+\,SDH\,+\,IVH\,+\,EPH (in order of anatomical distinctiveness), with previously introduced classes carried forward at each stage. This design isolates interference from newly introduced classes from model capacity effects, directly testing whether the framework exhibits bounded class-local degradation under explicit exclusivity constraints or catastrophic collapse under implicit class competition. We use the metrics defined in Section~\ref{ssec:eval_metrics}.

\subsubsection{Ablation I: Effect of components of the proposed framework}
\label{sssec:abl1}
We evaluate three ablation configurations on BraTS 2020 and BHSD datasets using the metrics described in Section~\ref{ssec:eval_metrics}: (i)~\textit{w/o binary guidance}: $\mathcal{F}_{\text{ag}}^{\text{bin}}$ is removed and multiclass maps $\mathcal{F}^{(l,c)}(x)$ bypass the modulation in Eq.~\eqref{eq:refinement}; (ii)~\textit{w/o aggregation network}: $\mathcal{A}(\cdot;\theta^c)$ is replaced by uniform averaging, discarding the learned scale weights $\tilde{w}^{(l,c)}$ in Eq.~\eqref{eq:aggmap}; and (iii)~\textit{w/o pseudo-label refinement}: the segmentation network is trained on raw thresholded maps $\hat{S}^{c}(p)$ from Eq.~\eqref{eq:classthresh}, omitting the morphological pseudo label refinement of Section~\ref{sec:postprocess}.

\subsubsection{Ablation II: Impact of loss function components}
\label{sssec:abl2}
We incrementally introduce loss components: ($\mathcal{L}_c$ $\to$ $\mathcal{L}_c + \mathcal{L}_{\text{sep}}$ $\to$ $\mathcal{L}_c + \mathcal{L}_{\text{sep}} + \mathcal{L}_{\text{agree}}$) on BraTS 2020 and BHSD datasets using the metrics described in Section~\ref{ssec:eval_metrics}, to quantify the individual contribution of each term and verify that each loss addresses a distinct and non-substitutable failure mode, as established in Section~\ref{sec:losses}.

\subsubsection{Ablation III: Analysis of multiscale resolution sensitivity}
\label{sssec:abl3}
We assess the importance of multi-scale aggregation by systematically retaining individual and pairs of adjacent CAM resolutions from the four-level multi-exit hierarchy (R1--R4, Fig.~\ref{fig:method}) using the metrics described in Section~\ref{ssec:eval_metrics}. We aim to quantify the effect of reduced scale diversity on segmentation quality across both datasets. 

\subsubsection{Uncertainty Analysis}
\label{sssec:uncertain}
Epistemic and aleatoric uncertainty are estimated on BraTS 2020 and BHSD to assess model reliability under clinical deployment conditions. Epistemic uncertainty is quantified via Monte Carlo Dropout~\citep{gal2016dropoutbayesianapproximationrepresenting}: dropout ($p = 0.15$) is injected after every ReLU activation of $\mathcal{A}(\cdot;\theta^c)$ at inference only, with $T = 30$ stochastic forward passes. Aleatoric uncertainty is estimated via Test-Time Augmentation~\citep{wang2019aleatoric} across $T = 12$ augmentations comprising horizontal and vertical flips, rotations of $\pm10^\circ$ and $\pm15^\circ$, and Gaussian noise ($\sigma = 0.5$). Three metrics are reported per class: Predictive Entropy $H = -\sum_c \bar{q}_c \log \bar{q}_c$, where $\hat{q}_{t,c}$ is the predicted class probability at pass $t$ and $\bar{q}_c = \frac{1}{T}\sum_t \hat{q}_{t,c}$ its mean; Mutual Information $\mathrm{MI} = H[\bar{q}] - \frac{1}{T}\sum_t H[\hat{q}_t]$, isolating the epistemic component; and Dice Uncertainty $\sigma(\{\mathrm{Dice}(\hat{q}_t > 0.4,\, y^{(c)})\}_{t=1}^T)$, quantifying segmentation variability in task-relevant units (reported as standard deviation across $T$ passes). Calibrated uncertainty estimates are a desirable property for clinical translation, potentially helping identify high-confidence failures where lesion subtype attribution may influence downstream decisions.

\subsubsection{Qualitative analysis of representations}
\label{sssec:qualitative}
We analyze t-SNE visualizations of class-discriminative activations for BraTS 2020, across three ablation axes, including incremental loss combination (Section~\ref{sssec:abl1}): $\mathcal{L}_c \to \mathcal{L}_c + \mathcal{L}_{\text{sep}} \to \mathcal{L}_c + \mathcal{L}_{\text{sep}} + \mathcal{L}_{\text{agree}}$, model component ablation (Section~\ref{sssec:abl1}): without binary guidance ($\mathcal{F}_{\text{ag}}^{\text{bin}}$) $\to$ $\mathcal{F}_{\text{ag}}^{\text{bin}}$ + without aggregation network $\to$ full model components, and resolution scale selection (Section~\ref{sssec:abl1}): single-scales alone $\to$ two-scales alone $\to$ full multi-exit hierarchy. We chose BraTS due to the higher degree of label adjacency compared to ICH datasets. 

\section{Results}
\label{sec:results}
\subsection{Comparison of encoder backbone architectures}
\label{ssec:arch_results}

\begin{figure*}[t]
\centering
\includegraphics[width=0.85\textwidth]{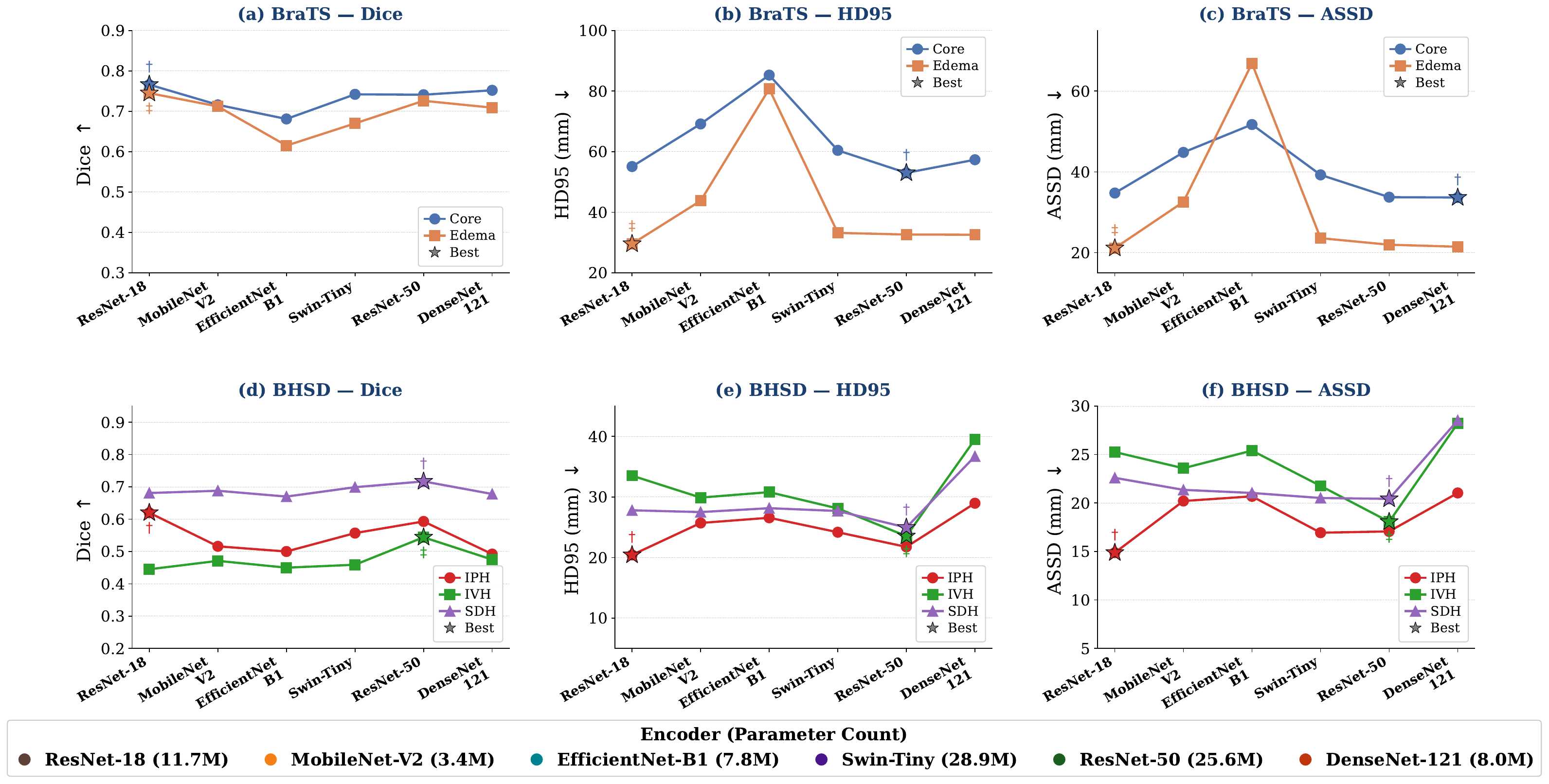}
\caption{Comparison of segmentation performance across six encoder backbone architectures on BraTS 2020 and BHSD datasets. Dice $\uparrow$, HD95 $\downarrow$ and ASSD $\downarrow$ values shown for Tumor Core and Edema (BraTS 2020) and IPH, IVH, and SDH (BHSD). $\bigstar$ indicates the best result per metric; $\dagger$ and $\ddagger$~indicate statistically significant improvement of the architecture over the second best and all other backbones respectively ($p < 0.05$, Wilcoxon signed-rank test).} 
\label{fig:backbone_analysis}
\end{figure*}

Fig.~\ref{fig:backbone_analysis} reports results of comparison across network architectures. Performance remains consistently strong across all encoders, validating that the observed gains are attributable to the binary-guided mutual exclusivity mechanism rather than backbone capacity. On BraTS, ResNet-18 achieves the highest Tumor Core Dice (0.766) and lowest Edema HD95 (29.56 mm), while ResNet-50 obtains the best Tumor Core HD95 (53.01 mm). EfficientNet-B1 exhibits the weakest boundary localization (Core HD95 $\approx$ 86 mm, Edema HD95 $\approx$ 82 mm), and Swin-Tiny, despite being the largest encoder, fails to improve upon ResNet-18 on either dataset, confirming that neither architectural efficiency nor raw model capacity is the primary driver of performance. On BHSD, ResNet-50 achieves the best Dice on IVH (0.54) and SDH (0.72) with the most stable HD95 across subtypes. While MobileNet-V2 also remains competitive on Dice, ResNet-50 achieves strictly superior boundary metrics across all three subtypes, indicating that its deeper feature hierarchy provides greater spatial precision for multiclass CT lesion discrimination. Backbone differences are statistically significant across all metrics ($p < 0.01$); ResNet-18 outperforms all encoders on Edema ($p < 0.01$) and ResNet-50 on IVH across all metrics ($p \leq 0.01$). Hence, ResNet-18 and ResNet-50 are chosen for further experiments on brain tumor and ICH respectively.

\subsection{Comparison with weakly supervised state-of-the-art methods}
\label{ssec:sota_results}
Table~\ref{tab:brats_results}, Table~\ref{tab:ich_results} and Fig.~\ref{fig:visual_res} report quantitative and qualitative state-of-the-art comparisons within BraTS 2020 and BHSD, with \textsf{BiMEx-MS} consistently outperforming all baselines with significance ($p < 0.01$).

\begin{table*}[t]
\centering
\caption{Comparison of segmentation results with those of state-of-the-art WSSS methods on the BraTS 2020 dataset (mean $\pm$ standard deviation). Dice, HD95 and ASSD values are reported. $\uparrow$/$\downarrow$ indicate higher/lower values are better. Best results are highlighted in \textbf{bold}, while the second-best results are \underline{underlined}. $^\dagger$ Statistically significant ($p < 0.01$, one-sided Wilcoxon signed-rank test) against top-performing baselines (MCTformer+, AME-CAM, FMA-WSSS).}
\label{tab:brats_results}
\footnotesize
\setlength{\tabcolsep}{3pt}
\renewcommand{\arraystretch}{1.05}
\begin{tabular}{lcccccc}
\toprule
& \multicolumn{3}{c}{\textbf{Tumor Core}} 
& \multicolumn{3}{c}{\textbf{Edema}} \\
\cmidrule(r){2-4} \cmidrule(l){5-7}
\textbf{Method} 
& Dice $\uparrow$ & HD95 (mm) $\downarrow$ & ASSD (mm) $\downarrow$
& Dice $\uparrow$ & HD95 (mm) $\downarrow$ & ASSD (mm) $\downarrow$ \\
\midrule
\multicolumn{7}{l}{\textit{a. Classical CAM-based methods}} \\[2pt]
GradCAM
& 0.623 $\pm$ 0.471 & 126.080 $\pm$ 173.34 & 99.169 $\pm$ 116.45
& 0.336 $\pm$ 0.444 & 170.295 $\pm$ 158.43 & 109.531 $\pm$ 93.15 \\
LayerCAM
& 0.617 $\pm$ 0.475 & 127.209 $\pm$ 172.91 & 98.360 $\pm$ 116.02
& 0.339 $\pm$ 0.441 & 164.188 $\pm$ 160.51 & 105.898 $\pm$ 91.93 \\
ScoreCAM
& 0.617 $\pm$ 0.475 & 127.970 $\pm$ 172.97 & 98.418 $\pm$ 116.02
& 0.337 $\pm$ 0.442 & 163.568 $\pm$ 160.30 & 105.060 $\pm$ 91.28 \\
EigenCAM
& 0.614 $\pm$ 0.442 & 127.765 $\pm$ 166.92 & 98.059 $\pm$ 116.09
& 0.336 $\pm$ 0.554 & 172.211 $\pm$ 161.64 & 114.588 $\pm$ 91.38 \\
ReCAM
& 0.637 $\pm$ 0.426 & 119.632 $\pm$ 122.33 & 87.235 $\pm$ 103.69
& 0.428 $\pm$ 0.459 & 149.713 $\pm$ 151.70 & 99.588 $\pm$ 91.43 \\
\midrule
\multicolumn{7}{l}{\textit{b. Medical WSSS methods}} \\[2pt]
AME-CAM
& 0.698 $\pm$ 0.426 & \underline{59.713} $\pm$ 112.49 & 49.530 $\pm$ 110.80
& \underline{0.652} $\pm$ 0.401 & \underline{54.207} $\pm$ 97.66  & \underline{38.378} $\pm$ 96.69 \\
HAMIL
& 0.552 $\pm$ 0.381 & 89.642 $\pm$ 111.42 & 61.428 $\pm$ 107.24
& 0.488 $\pm$ 0.438 & 94.927 $\pm$ 109.71 & 73.668 $\pm$ 94.41 \\
PathMamba
& 0.568 $\pm$ 0.412 & 84.426 $\pm$ 111.72 & 58.922 $\pm$ 100.42
& 0.404 $\pm$ 0.471 & 104.924 $\pm$ 121.62 & 79.429 $\pm$ 99.62 \\
ESFAN
& 0.646 $\pm$ 0.398 & 89.027 $\pm$ 102.11 & 63.530 $\pm$ 113.60
& 0.522 $\pm$ 0.487 & 101.774 $\pm$ 137.53 & 78.378 $\pm$ 89.44 \\
\midrule
\multicolumn{7}{l}{\textit{c. Transformer-based methods}} \\[2pt]
MCTformer
& 0.664 $\pm$ 0.328 & 87.286 $\pm$ 103.46 & 57.428 $\pm$ 106.25
& 0.606 $\pm$ 0.388 & 80.728 $\pm$ 98.09  & 48.349 $\pm$ 99.13 \\
Sparse-ViT
& 0.664 $\pm$ 0.394 & 81.416 $\pm$ 103.46 & 55.238 $\pm$ 98.03
& 0.515 $\pm$ 0.417 & 89.916 $\pm$ 102.42 & 66.652 $\pm$ 97.25 \\
MCTformer+
& \underline{0.754} $\pm$ 0.378 & 63.503 $\pm$ 81.73 & \underline{39.242} $\pm$ 75.76
& 0.652 $\pm$ 0.421 & 69.006 $\pm$ 77.53 & 41.204 $\pm$ 71.86 \\
CoSA
& 0.677 $\pm$ 0.402 & 84.916 $\pm$ 102.42 & 59.632 $\pm$ 98.31
& 0.476 $\pm$ 0.412 & 98.427 $\pm$ 116.82 & 71.529 $\pm$ 95.85 \\
WeakTR
& 0.524 $\pm$ 0.392 & 76.698 $\pm$ 95.84  & 56.888 $\pm$ 65.01
& 0.381 $\pm$ 0.341 & 84.965 $\pm$ 95.93  & 63.189 $\pm$ 50.89 \\
\midrule
\multicolumn{7}{l}{\textit{d. Foundation model-assisted methods}} \\[2pt]
FMA-WSSS
& 0.693 $\pm$ 0.432 & 82.576 $\pm$ 100.89 & 52.778 $\pm$ 80.38
& 0.636 $\pm$ 0.416 & 69.450 $\pm$ 103.40 & 38.107 $\pm$ 34.63 \\
SemPLeS
& 0.710 $\pm$ 0.424 & 70.168 $\pm$ 95.56  & 45.596 $\pm$ 84.37
& 0.614 $\pm$ 0.462 & 73.025 $\pm$ 93.06  & 47.506 $\pm$ 85.12 \\
\midrule
\textbf{\textsf{BiMEx-MS} (ResNet-18)}
& \textbf{0.766} $\pm$ 0.292 & 55.093 $\pm$ 63.06          & 34.757 $\pm$ 50.65
& \textbf{0.745}$^\dagger$ $\pm$ 0.251 & \textbf{29.558}$^\dagger$ $\pm$ 44.64 & \textbf{21.128}$^\dagger$ $\pm$ 34.06 \\
\textbf{\textsf{BiMEx-MS} (ResNet-50)}
& 0.741 $\pm$ 0.282 & \textbf{53.008}$^\dagger$ $\pm$ 59.63 & \textbf{33.733} $\pm$ 50.60
& 0.726 $\pm$ 0.276 & 32.644 $\pm$ 40.93          & 21.964 $\pm$ 33.95 \\
\bottomrule
\end{tabular}
\end{table*}

Classical CAM-based methods establish a performance floor on both benchmarks: Tumor Core Dice of 0.614--0.623 but Edema Dice of only 0.336--0.339 with HD95 exceeding 160 mm on BraTS (Table~\ref{tab:brats_results}), and within 0.02 Dice of one another across all BHSD subtypes (Table~\ref{tab:ich_results}). ReCAM's softmax reactivation improves marginally (Tumor Core 0.637, Edema 0.428) but retains background leakage without explicit confinement to $\Omega^+$. Without spatial and inter-class constraints, gradient-based methods converge to overlapping activations regardless of saliency definition (Fig.~\ref{fig:visual_res}).

\begin{table*}[!tbp]
\centering
\caption{Comparison of segmentation results with those of state-of-the-art WSSS methods on the BHSD dataset (mean $\pm$ standard deviation). Dice, HD95 and ASSD values are reported. $\uparrow$/$\downarrow$ indicate higher/lower values are better. Best results are highlighted in \textbf{bold}, while the second-best results are \underline{underlined}. $^\dagger$ Statistically significant ($p < 0.01$, one-sided Wilcoxon signed-rank test) against top-performing baselines (MCTformer+, AME-CAM, FMA-WSSS).}
\label{tab:ich_results}
\footnotesize
\setlength{\tabcolsep}{3pt}
\renewcommand{\arraystretch}{1}
\begin{tabular}{lcccccc}
\toprule
& \multicolumn{3}{c}{\textbf{EPH}} 
& \multicolumn{3}{c}{\textbf{IPH}}  \\
\cmidrule(r){2-4} \cmidrule(lr){5-7} 
\textbf{Method} 
& Dice $\uparrow$ & HD95 (mm) $\downarrow$ & ASSD (mm) $\downarrow$
& Dice $\uparrow$ & HD95 (mm) $\downarrow$ & ASSD (mm) $\downarrow$\\
\midrule
\multicolumn{7}{l}{\textit{a. Classical CAM-based methods}} \\[2pt]
GradCAM
& 0.442 $\pm$ 0.40 & 35.02 $\pm$ 50.08  & 29.25 $\pm$ 50.75
& 0.468 $\pm$ 0.43 & 31.52 $\pm$ 50.93  & 21.58 $\pm$ 51.55\\
LayerCAM
& 0.486 $\pm$ 0.41 & 32.32 $\pm$ 50.96  & 26.39 $\pm$ 50.31
& 0.432 $\pm$ 0.46 & 29.53 $\pm$ 50.69  & 22.54 $\pm$ 51.53\\
ScoreCAM
& 0.455 $\pm$ 0.43 & 35.32 $\pm$ 51.22  & 27.20 $\pm$ 49.80
& 0.452 $\pm$ 0.45 & 31.04 $\pm$ 50.69  & 21.45 $\pm$ 52.51\\
EigenCAM
& 0.484 $\pm$ 0.41 & 32.20 $\pm$ 50.92  & 25.02 $\pm$ 50.35
& 0.421 $\pm$ 0.47 & 29.36 $\pm$ 50.72  & 23.24 $\pm$ 51.44\\
ReCAM
& 0.490 $\pm$ 0.40 & 32.48 $\pm$ 49.03  & 24.56 $\pm$ 50.26
& 0.452 $\pm$ 0.44 & 29.64 $\pm$ 50.66  & 21.66 $\pm$ 51.59\\
\midrule
\multicolumn{7}{l}{\textit{b. Medical WSSS methods}} \\[2pt]
AME-CAM
& \underline{0.583} $\pm$ 0.34 & 26.33 $\pm$ 50.32  & 20.54 $\pm$ 49.49
& 0.536 $\pm$ 0.36 & 27.61 $\pm$ 47.13  & \underline{20.63} $\pm$ 50.25\\
HAMIL
& 0.487 $\pm$ 0.43 & 32.20 $\pm$ 45.82  & 25.01 $\pm$ 45.25
& 0.426 $\pm$ 0.46 & 29.40 $\pm$ 50.71  & 22.92 $\pm$ 51.48\\
PathMamba
& 0.471 $\pm$ 0.40 & 32.91 $\pm$ 46.17  & 25.47 $\pm$ 48.74
& 0.426 $\pm$ 0.48 & 29.12 $\pm$ 50.70  & 23.43 $\pm$ 51.55\\
ESFAN
& 0.479 $\pm$ 0.41 & 32.67 $\pm$ 46.35  & 25.31 $\pm$ 48.43
& 0.443 $\pm$ 0.42 & 27.16 $\pm$ 50.73  & 21.25 $\pm$ 51.75\\
\midrule
\multicolumn{7}{l}{\textit{c. Transformer-based methods}} \\[2pt]
MCTformer
& 0.517 $\pm$ 0.40 & 28.52 $\pm$ 50.88  & 21.36 $\pm$ 46.12
& 0.460 $\pm$ 0.44 & 32.50 $\pm$ 50.63  & 21.44 $\pm$ 51.50\\
Sparse-ViT
& 0.485 $\pm$ 0.40 & 32.86 $\pm$ 46.30  & 25.45 $\pm$ 48.44
& 0.457 $\pm$ 0.42 & 27.08 $\pm$ 50.70  & 20.69 $\pm$ 51.72\\
MCTformer+
& 0.529 $\pm$ 0.40 & 28.99 $\pm$ 50.50  & 21.34 $\pm$ 46.14
& 0.468 $\pm$ 0.44 & 32.54 $\pm$ 50.87  & 21.53 $\pm$ 51.65\\
CoSA
& 0.476 $\pm$ 0.40 & 33.40 $\pm$ 46.40  & 25.03 $\pm$ 48.32
& 0.469 $\pm$ 0.41 & 30.27 $\pm$ 51.21  & 21.35 $\pm$ 51.27\\
WeakTR
& 0.542 $\pm$ 0.39 & 27.40 $\pm$ 50.10  & 21.13 $\pm$ 45.66
& 0.468 $\pm$ 0.43 & 31.65 $\pm$ 50.92  & 21.65 $\pm$ 51.54\\
\midrule
\multicolumn{7}{l}{\textit{d. Foundation model-assisted methods}} \\[2pt]
FMA-WSSS
& 0.571 $\pm$ 0.35 & \underline{25.56} $\pm$ 50.05  & \underline{20.27} $\pm$ 48.69
& \underline{0.550} $\pm$ 0.35 & \underline{27.61} $\pm$ 46.71  & 21.35 $\pm$ 50.14\\
SemPLeS
& 0.508 $\pm$ 0.40 & 28.39 $\pm$ 50.76  & 20.95 $\pm$ 45.52
& 0.460 $\pm$ 0.44 & 32.57 $\pm$ 50.70  & 21.92 $\pm$ 51.95\\
\midrule
\textbf{\textsf{BiMEx-MS} (ResNet-18)}
& 0.581 $\pm$ 0.31          & 25.08 $\pm$ 50.06          & 20.43 $\pm$ 49.83
& 0.618 $\pm$ 0.31 & \textbf{21.45}$^\dagger$ $\pm$ 49.83 & \textbf{15.77}$^\dagger$ $\pm$ 47.35\\
\textbf{\textsf{BiMEx-MS} (ResNet-50)}
& \textbf{0.592}$^\dagger$ $\pm$ 0.33 & \textbf{24.38}$^\dagger$ $\pm$ 48.32 & \textbf{19.81} $\pm$ 47.59
& \textbf{0.631} $\pm$ 0.40          & 22.25 $\pm$ 52.79         & 17.94 $\pm$ 49.39\\
\bottomrule
\toprule
& \multicolumn{3}{c}{\textbf{IVH}} 
& \multicolumn{3}{c}{\textbf{SDH}} \\
\cmidrule(r){2-4} \cmidrule(lr){5-7} 
\textbf{Method} 
& Dice $\uparrow$ & HD95 (mm) $\downarrow$ & ASSD (mm) $\downarrow$
& Dice $\uparrow$ & HD95 (mm) $\downarrow$ & ASSD (mm) $\downarrow$\\
\midrule
\multicolumn{7}{l}{\textit{a. Classical CAM-based methods}} \\[2pt]
GradCAM
& 0.470 $\pm$ 0.49 & 36.07 $\pm$ 42.92  & 20.22 $\pm$ 36.13
& 0.548 $\pm$ 0.46 & 34.47 $\pm$ 58.54  & 29.21 $\pm$ 58.66\\
LayerCAM
& 0.479 $\pm$ 0.48 & 38.56 $\pm$ 46.34  & 23.18 $\pm$ 45.65
& 0.528 $\pm$ 0.48 & 33.67 $\pm$ 58.08  & 26.72 $\pm$ 58.17\\
ScoreCAM
& 0.437 $\pm$ 0.48 & 37.21 $\pm$ 47.27  & 20.51 $\pm$ 40.91
& 0.535 $\pm$ 0.49 & 33.73 $\pm$ 55.92  & 26.53 $\pm$ 56.94\\
EigenCAM
& 0.478 $\pm$ 0.48 & 38.41 $\pm$ 46.31  & 23.26 $\pm$ 45.18
& 0.527 $\pm$ 0.48 & 33.73 $\pm$ 58.07  & 25.84 $\pm$ 58.15\\
ReCAM
& 0.480 $\pm$ 0.48 & 38.57 $\pm$ 46.33  & 24.58 $\pm$ 40.51
& 0.528 $\pm$ 0.48 & 33.61 $\pm$ 58.09  & 25.57 $\pm$ 58.19\\
\midrule
\multicolumn{7}{l}{\textit{b. Medical WSSS methods}} \\[2pt]
AME-CAM
& \underline{0.498} $\pm$ 0.46 & \underline{26.50} $\pm$ 43.51  & \underline{20.08} $\pm$ 41.56
& 0.592 $\pm$ 0.43 & 31.74 $\pm$ 62.77  & 25.29 $\pm$ 51.24\\
HAMIL
& 0.479 $\pm$ 0.48 & 37.91 $\pm$ 46.31  & 23.16 $\pm$ 40.55
& 0.527 $\pm$ 0.48 & 33.73 $\pm$ 58.08  & 25.82 $\pm$ 58.15\\
PathMamba
& 0.477 $\pm$ 0.48 & 37.52 $\pm$ 45.96  & 23.03 $\pm$ 41.09
& 0.510 $\pm$ 0.46 & 33.76 $\pm$ 57.42  & 25.71 $\pm$ 58.29\\
ESFAN
& 0.446 $\pm$ 0.48 & 39.95 $\pm$ 46.37  & 23.67 $\pm$ 41.56
& 0.536 $\pm$ 0.47 & 32.60 $\pm$ 56.23  & 25.29 $\pm$ 58.63\\
\midrule
\multicolumn{7}{l}{\textit{c. Transformer-based methods}} \\[2pt]
MCTformer
& 0.458 $\pm$ 0.49 & 36.75 $\pm$ 45.08  & 20.82 $\pm$ 36.39
& 0.533 $\pm$ 0.46 & 34.20 $\pm$ 57.76  & 27.91 $\pm$ 59.04\\
Sparse-ViT
& 0.474 $\pm$ 0.49 & 39.96 $\pm$ 46.62  & 23.80 $\pm$ 41.45
& 0.574 $\pm$ 0.47 & 33.67 $\pm$ 56.39  & 25.73 $\pm$ 58.23\\
MCTformer+
& 0.469 $\pm$ 0.50 & 37.16 $\pm$ 44.99  & 21.14 $\pm$ 36.71
& 0.530 $\pm$ 0.44 & 34.87 $\pm$ 57.42  & 28.03 $\pm$ 58.66\\
CoSA
& 0.466 $\pm$ 0.49 & 40.77 $\pm$ 46.14  & 23.96 $\pm$ 41.61
& 0.551 $\pm$ 0.44 & 33.75 $\pm$ 56.30  & 25.63 $\pm$ 58.01\\
WeakTR
& 0.470 $\pm$ 0.48 & 36.64 $\pm$ 42.98  & 20.45 $\pm$ 36.17
& 0.548 $\pm$ 0.45 & 34.53 $\pm$ 58.55  & 28.22 $\pm$ 58.66\\
\midrule
\multicolumn{7}{l}{\textit{d. Foundation model-assisted methods}} \\[2pt]
FMA-WSSS
& 0.487 $\pm$ 0.41 & 26.95 $\pm$ 43.53  & 20.54 $\pm$ 41.70
& \underline{0.608} $\pm$ 0.45 & \underline{31.67} $\pm$ 60.21  & \underline{22.58} $\pm$ 51.32\\
SemPLeS
& 0.472 $\pm$ 0.48 & 38.93 $\pm$ 43.08  & 22.08 $\pm$ 35.82
& 0.541 $\pm$ 0.47 & 34.68 $\pm$ 57.42  & 27.25 $\pm$ 59.03\\
\midrule
\textbf{\textsf{BiMEx-MS} (ResNet-18)}
& 0.535 $\pm$ 0.44          & 30.51 $\pm$ 47.33          & 21.79 $\pm$ 45.11
& 0.641 $\pm$ 0.41          & \textbf{30.39}$^\dagger$ $\pm$ 62.91         & \textbf{22.29} $\pm$ 53.16\\
\textbf{\textsf{BiMEx-MS} (ResNet-50)}
& \textbf{0.593}$^\dagger$ $\pm$ 0.48 & \textbf{24.11} $\pm$ 35.97  & \textbf{18.56} $\pm$ 38.03
& \textbf{0.664}$^\dagger$ $\pm$ 0.42 & 30.94 $\pm$ 56.42  & 25.15 $\pm$ 52.58\\
\bottomrule
\end{tabular}%
\end{table*}

\textsf{BiMEx-MS} achieves the best performance on both datasets (Tables~\ref{tab:brats_results} and~\ref{tab:ich_results}). On BraTS, ResNet-18 attains Tumor Core Dice 0.766 and Edema Dice 0.745 with Edema HD95 29.558 mm, the only method below 40 mm across all compared approaches, while ResNet-50 achieves the best Tumor Core HD95 and ASSD (53.008 mm and 33.733 mm). Improvements over AME-CAM and MCTformer+ are statistically significant on Edema Dice ($p < 0.001$), as are boundary metric gains over FMA-WSSS for IPH ($p = 0.047$), SDH ($p < 0.01$), and all BraTS metrics ($p < 0.001$). On BHSD, ResNet-50 leads all subtypes, with gains on SDH (Dice 0.664) and IVH (Dice 0.593, HD95 24.11 mm vs.\ 26.95 mm for FMA-WSSS). Improvements are consistently larger on boundary metrics and rarer classes, reflecting the orthogonality enforced by $\mathcal{L}_{\text{sep}}$ and the frequency-agnostic coverage of $\mathcal{F}_{\text{ag}}^{\text{bin}}$.

Among competing methods, AME-CAM is the strongest baseline (Edema Dice 0.652, Tumor Core HD95 59.713 mm on BraTS; IVH Dice 0.498, EPH Dice 0.583 on BHSD), however its class-agnostic aggregation produces diffuse activations that spread into neighboring territories. HAMIL, PathMamba, and ESFAN perform below AME-CAM on both datasets, as histopathology-designed methods fail to resolve gradual intensity transitions in neuroimaging. MCTformer+ achieves the second-best Tumor Core Dice (0.754, Table~\ref{tab:brats_results}) but offers no advantage over classical CAMs on BHSD (EPH 0.529, IVH 0.469; Table~\ref{tab:ich_results}), confirming that attention-based separation does not transfer to CT. FMA-WSSS performs competitively on compact lesions (IPH 0.550, SDH 0.608) but degrades on diffuse subtypes (IVH 0.487) and exhibits subtype misassignment between co-occurring classes (Fig.~\ref{fig:visual_res}(d)).

\subsection{Cross-Dataset Generalization}
\label{ssec:crossdata_results}

\begin{table}[p]
\centering
\caption{Segmentation results for cross-dataset generalization. Source-trained models evaluated on target datasets without fine-tuning. $^\dagger$Results on full BHSD (192 volumes), reported as reference.}
\label{tab:cross_dataset}
\scriptsize
\setlength{\tabcolsep}{3.5pt}
\renewcommand{\arraystretch}{1.0}
\begin{tabular}{llccc}
\toprule
\textbf{Dataset} & \textbf{Class} & \textbf{Dice} $\uparrow$ & \textbf{HD95 (mm)} $\downarrow$ & \textbf{ASSD (mm)} $\downarrow$ \\
\midrule
\multirow{2}{*}{\parbox{2.2cm}{BraTS 2020$\to$\\BraTS 2023 SSA}}
& Core  & $0.759 \pm 0.36$ & $57.84 \pm 97.60$  & $44.72 \pm 98.19$ \\
& Edema & $0.711 \pm 0.36$ & $50.84 \pm 107.36$ & $39.68 \pm 98.43$ \\
\midrule
\multirow{4}{*}{\parbox{2.2cm}{RSNA-ICH$\to$\\BHSD (Test split)}}
& EPH & $0.562 \pm 0.34$ & $24.49 \pm 50.70$  & $20.28 \pm 49.74$ \\
& IPH & $0.598 \pm 0.34$ & $26.39 \pm 51.05$  & $20.53 \pm 50.56$ \\
& IVH & $0.582 \pm 0.40$ & $24.52 \pm 38.23$  & $19.42 \pm 37.34$ \\
& SDH & $0.660 \pm 0.41$ & $30.13 \pm 56.85$  & $25.13 \pm 51.05$ \\
\midrule
\multirow{4}{*}{\parbox{2.2cm}{RSNA-ICH$\to$\\BHSD (Full$^\dagger$)}}
& EPH & $0.623 \pm 0.34$ & $23.37 \pm 50.26$  & $19.68 \pm 49.13$ \\
& IPH & $0.615 \pm 0.35$ & $23.70 \pm 52.25$  & $18.30 \pm 47.36$ \\
& IVH & $0.563 \pm 0.42$ & $24.94 \pm 37.69$  & $18.56 \pm 35.77$ \\
& SDH & $0.704 \pm 0.42$ & $27.19 \pm 58.71$  & $22.74 \pm 48.50$ \\
\bottomrule
\end{tabular}
\end{table}

Table~\ref{tab:cross_dataset} reports cross-dataset generalization performance. On BraTS 2023 SSA, the BraTS 2020-trained model achieves Core Dice 0.759 and Edema Dice 0.711 without fine-tuning, closely approaching within-dataset results (Core 0.766, Edema 0.745 in Table~\ref{tab:brats_results}), confirming robust cross-dataset transfer. For ICH, the RSNA-trained model yields Dice values of 0.562, 0.598, 0.582, and 0.660 for EPH, IPH, IVH, and SDH respectively on BHSD, marginally below in-domain performance (except for IVH) with none of the differences statistically significant. Also, EPH Dice is comparable to other classes indicating good cross-data transfer on the tail class. The binary prior $\mathcal{F}_{\text{ag}}^{\text{bin}}$ specifically contributes to cross-domain robustness by grounding localization in global image features that generalize more reliably than fine-level subtype-specific activations.

\subsection{Evaluation under Progressive Class Expansion}
\label{ssec:class_exp_results}
Table~\ref{tab:class_progression} reports progressive label expansion results on the RSNA$\,\to\,$BHSD pipeline using ResNet-50. At two classes, the framework separates IPH (Dice 0.673, HD95 21.18 mm) and SDH (Dice 0.638, HD95 31.25 mm) cleanly. Introducing IVH reduces IPH Dice to 0.593 while improving SDH Dice to 0.717 (HD95 24.96 mm), attributable to $\mathcal{L}_{\text{sep}}$ orthogonalizing intraventricular activations away from the subdural channel. Adding EPH as the fourth and rarest class achieves EPH Dice 0.623 (HD95 23.37 mm), with stable values of existing metrics: HD95 variations below 3 mm and maximum Dice change of 0.022 across pre-existing classes, confirming that mutual exclusivity enforced by $\mathcal{F}_{\text{ag}}^{\text{bin}}$ and $\mathcal{L}_{\text{sep}}$ scales gracefully to four classes without catastrophic degradation.
Notably, on EPH, which is the tail class, we achieve competitive Dice (0.623) without any degradation to pre-existing classes, directly validating the class-frequency-agnostic coverage enforced by $\mathcal{F}_{\text{ag}}^{\text{bin}}$.

\begin{table}[b]
\centering
\caption{Segmentation performance of \textsf{BiMEx-MS} under progressive class expansion on the BHSD dataset using ResNet-50 backbone (mean $\pm$ standard deviation), along with \% variation in Dice. The newly added class highlighted in \textbf{bold}.}
\label{tab:class_progression}
\scriptsize
\setlength{\tabcolsep}{3pt}
\begin{tabular}{l l cccc}
\toprule
\textbf{Setting} & \textbf{Class} & \textbf{Dice\,$\uparrow$} & \textbf{HD95 (mm)\,$\downarrow$} & \textbf{ASSD (mm)\,$\downarrow$}  & \textbf{$\Delta$ Dice (\%) } \\
\midrule
\multirow{2}{*}{2-class}
 & IPH & $0.673 \pm 0.38$ & $21.18 \pm 50.76$  & $16.86 \pm 50.58$ & --- \\
 & SDH & $0.638 \pm 0.41$ & $31.25 \pm 57.70$  & $26.25 \pm 55.89$ & --- \\
\midrule
\multirow{3}{*}{3-class}
 & IPH & $0.593 \pm 0.35$ & $21.76 \pm 51.03$  & $17.07 \pm 50.30$ & $-11.9\%$ \\
 & SDH & $0.717 \pm 0.40$ & $24.96 \pm 56.75$  & $20.43 \pm 52.30$ & $+12.4\%$ \\
 & \textbf{IVH} & $0.544 \pm 0.42$ & $23.54 \pm 48.87$  & $18.07 \pm 44.89$ & --- \\
\midrule
\multirow{4}{*}{4-class}
 & IPH & $0.615 \pm 0.35$ & $23.70 \pm 52.25$  & $18.30 \pm 47.36$ & $+3.7\%$ \\
 & SDH & $0.704 \pm 0.42$ & $27.19 \pm 58.71$  & $22.74 \pm 48.50$ & $-1.8\%$ \\
 & IVH & $0.563 \pm 0.42$ & $24.94 \pm 37.69$  & $18.56 \pm 35.77$ & $+3.5\%$ \\
 & \textbf{EPH} & $0.623 \pm 0.34$ & $23.37 \pm 50.26$  & $19.68 \pm 49.13$ & --- \\
\bottomrule
\end{tabular}
\end{table}

\begin{figure*}[p]
\centering
\includegraphics[width=1.00\textwidth]{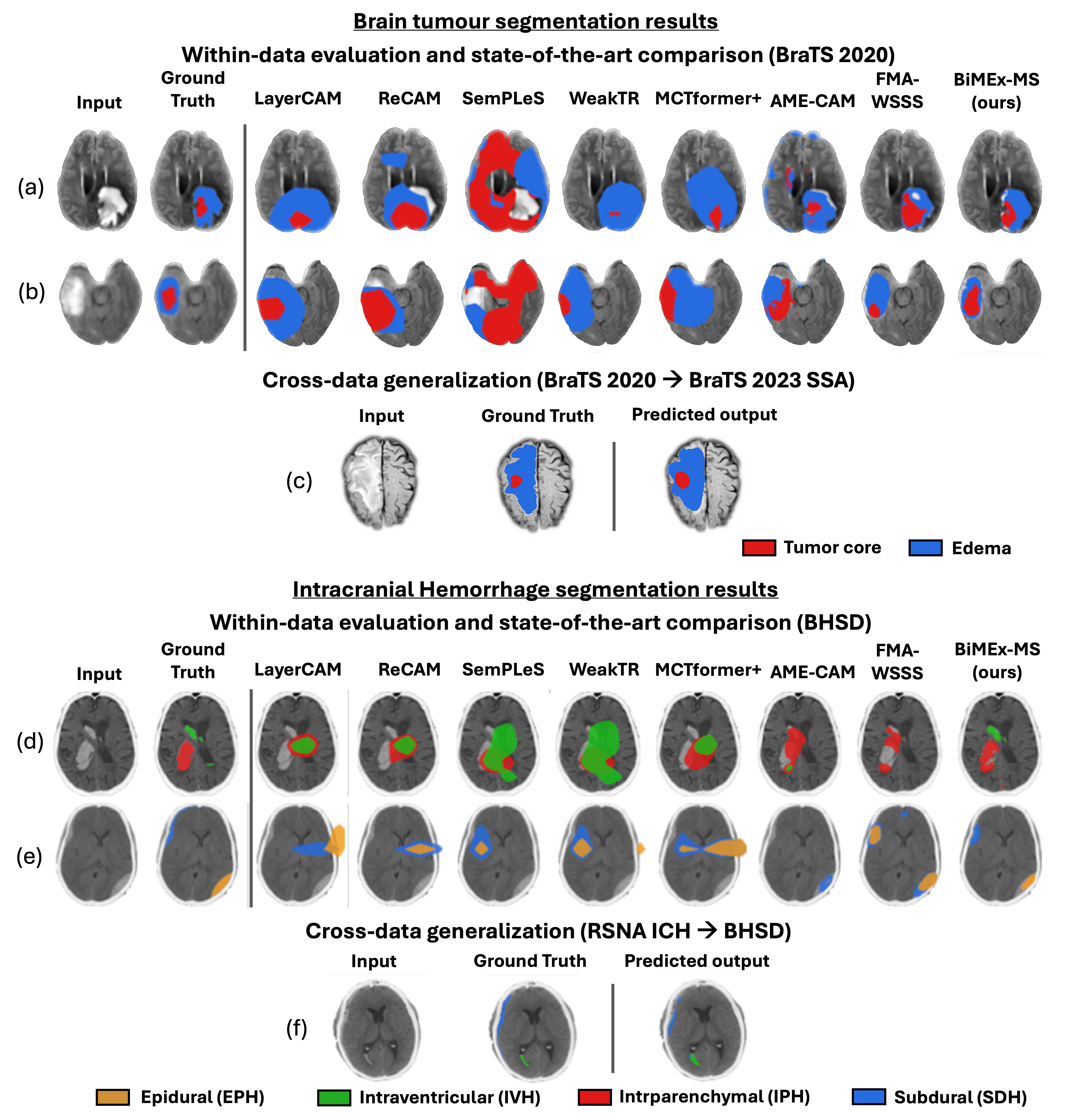}
\caption{Sample comparison results of \textsf{BiMEx-MS} with state-of-the-art WSSS methods on the full axial field of view of brain slices. Top panel: within-dataset evaluation on BraTS~2020 (a,b) and cross-continental generalization on BraTS~2020$\to$BraTS~2023~SSA (c) for core (red) and edema (blue) classes. Bottom panel: within-dataset evaluation on BHSD (d,e) and cross-dataset generalization for RSNA$\to$BHSD (f) for EPH (dark yellow), IVH (green), IPH (red) and SDH (blue) classes. \textsf{BiMEx-MS} yields sharp, mutually exclusive activation maps similar to the ground truth.} 
\label{fig:visual_res}
\end{figure*}

\begin{figure*}[t]
\centering
\includegraphics[width=0.85\textwidth]{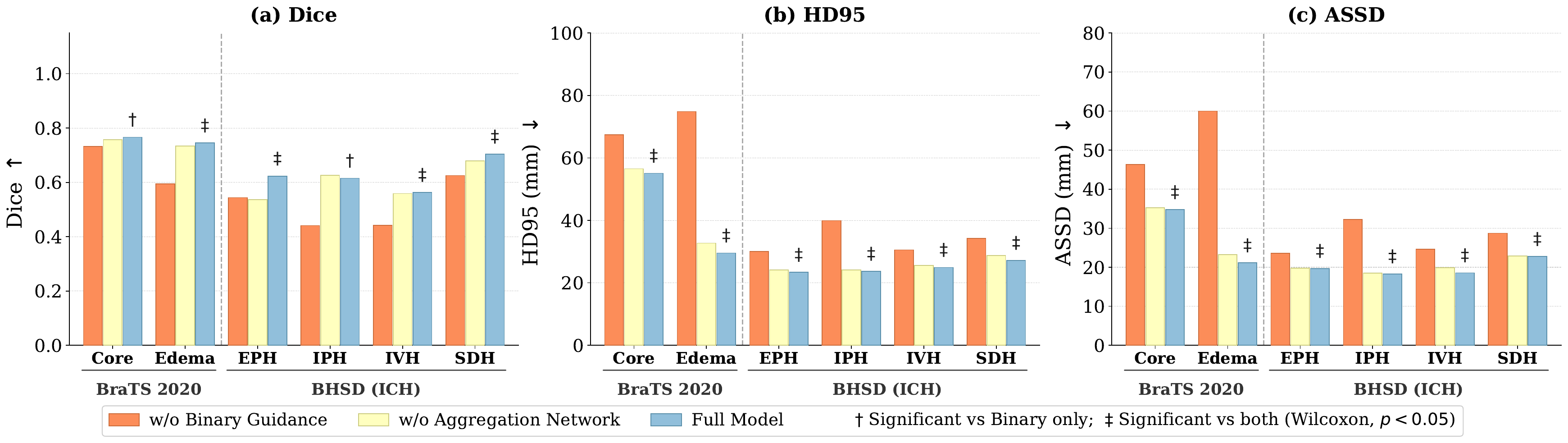}
\caption{Ablation results showing the effect of framework components on BraTS 2020 and BHSD datasets. Dice $\uparrow$ (a), HD95 $\downarrow$ (b) and ASSD $\downarrow$ (c) shown for the settings: (orange) without binary guidance from $\mathcal{F}_{\text{ag}}^{\text{bin}}$, (yellow) replacement of the class-specific aggregation network with uniform averaging and (blue) the complete \textsf{BiMEx-MS} framework. Removal of binary guidance causes the most severe degradation, particularly for rare subtypes (IVH and EPH) on BHSD, confirming that $\mathcal{F}_{\text{ag}}^{\text{bin}}$ provides frequency-agnostic structural coverage even for the most underrepresented subtype.}
\label{fig:component_ablation}
\end{figure*}

\subsection{Ablation I: Effect of Framework Components}
\label{ssec:abl1_results}

Fig.~\ref{fig:component_ablation} summarizes the contribution of each framework component across BraTS and BHSD. Removing binary guidance causes the most severe degradation: Edema Dice drops from 0.745 to 0.590 on BraTS and IPH and IVH Dice collapse to 0.44 on BHSD, confirming that $\mathcal{F}_{\text{ag}}^{\text{bin}}$ is the primary structural constraint preventing inter-class leakage. Replacing the class-specific aggregation network with uniform scale averaging yields a moderate drop, where tumor Core Dice falls from 0.766 to 0.760 on BraTS and EPH Dice from 0.623 to 0.540 on BHSD, demonstrating that class-conditioned scale weighting contributes meaningful discriminability beyond binary guidance alone. Omitting morphological pseudo-label refinement reduces Tumor Core Dice from 0.766 to 0.748 and Edema Dice from 0.745 to 0.736 on BraTS, with consistent degradation on BHSD, confirming that hierarchical refinement enforces a degree of pixel-wise mutual exclusivity that CAM-level outputs alone cannot guarantee. All ablation differences are statistically significant ($p < 0.01$), with HD95 degradation consistently larger than Dice, confirming the role of each component in spatial precision beyond the overall overlap.

\subsection{Ablation II: Impact of Loss Function Components}
\label{sssec:loss_abl}

Table~\ref{tab:loss_ablation} reports the improvements in performance with the incremental addition of loss terms. The $\mathcal{L}_c$ component alone yields competent coarse segmentation (BraTS: Tumor Core Dice 0.739, Edema 0.716; BHSD: IPH 0.622, SDH 0.678), confirming that per-class foreground--background separation is a necessary foundation. Adding $\mathcal{L}_{\text{sep}}$ improves inter-class spatial separability, where EPH Dice rises to 0.608 and IVH to 0.541 on BHSD, but transiently reduces IPH Dice to 0.576 and increases BraTS Tumor Core HD95 to 60.11 mm, reflecting a tension between feature orthogonality and class-wise coverage prior to spatial grounding. The full objective $\mathcal{L}_c + \mathcal{L}_{\text{sep}} + \mathcal{L}_{\text{agree}}$ resolves this tension: $\mathcal{L}_{\text{agree}}$ anchors multiclass predictions to the binary lesion prior, recovering Dice across all classes and improving boundary metrics. The sole exception is IPH, where the full loss yields marginally lower Dice (0.615 vs.\ 0.622) and higher HD95 relative to $\mathcal{L}_c$ alone, a minor trade-off attributable to increased inter-class competition from co-occurring subtypes in the cross-dataset setting. 

\begin{table*}[b]
\centering
\caption{Ablation study on loss function components on BraTS 2020 and BHSD datasets. $\uparrow$/$\downarrow$ indicate higher/lower values are better. Results are reported using the best-performing backbone per dataset (BraTS 2020: ResNet-18, BHSD: ResNet-50). Best results are shown in \textbf{bold}. $^\dagger$ Statistically significant compared to other settings ($p < 0.01$, one-sided Wilcoxon signed-rank test).}
\label{tab:loss_ablation}
\setlength{\tabcolsep}{4pt}
\renewcommand{\arraystretch}{1.05}
\resizebox{\textwidth}{!}{%
\begin{tabular}{lcccccccccccc}
\toprule
& \multicolumn{4}{c}{\textbf{BraTS 2020}} 
& \multicolumn{8}{c}{\textbf{ICH (RSNA $\to$ BHSD)}} \\
\cmidrule(r){2-5} \cmidrule(l){6-13}
& \multicolumn{2}{c}{\textbf{Core}} 
& \multicolumn{2}{c}{\textbf{Edema}}
& \multicolumn{2}{c}{\textbf{EPH}} 
& \multicolumn{2}{c}{\textbf{IPH}} 
& \multicolumn{2}{c}{\textbf{IVH}} 
& \multicolumn{2}{c}{\textbf{SDH}} \\
\cmidrule(r){2-3} \cmidrule(r){4-5} 
\cmidrule(r){6-7} \cmidrule(r){8-9} 
\cmidrule(r){10-11} \cmidrule(l){12-13}
\textbf{Loss Configuration}
& Dice $\uparrow$ & HD95 (mm) $\downarrow$
& Dice $\uparrow$ & HD95 (mm) $\downarrow$
& Dice $\uparrow$ & HD95 (mm) $\downarrow$
& Dice $\uparrow$ & HD95 (mm) $\downarrow$
& Dice $\uparrow$ & HD95 (mm) $\downarrow$
& Dice $\uparrow$ & HD95 (mm) $\downarrow$ \\
\midrule
$\mathcal{L}_c$
& 0.739 & \textbf{54.47}
& 0.716 & 32.25
& 0.596 & 27.46
& \textbf{0.622} & \textbf{22.36}
& 0.530 & 25.64
& 0.678 & 28.01 \\
$\mathcal{L}_c + \mathcal{L}_{\text{sep}}$
& 0.745 & 60.11
& 0.698 & 32.56
& 0.608 & 23.67
& 0.576 & 23.72
& 0.541 & \textbf{23.40}
& 0.692 & 27.27 \\
$\mathcal{L}_c + \mathcal{L}_{\text{sep}} + \mathcal{L}_{\text{agree}}$
& \textbf{0.766}$^\dagger$ & 55.09
& \textbf{0.745}$^\dagger$ & \textbf{29.56}$^\dagger$
& \textbf{0.623}$^\dagger$ & \textbf{23.37}
& 0.615 & 23.70
& \textbf{0.563}$^\dagger$ & 24.94
& \textbf{0.704}$^\dagger$ & \textbf{27.19} \\
\bottomrule
\end{tabular}%
}
\end{table*}

\begin{figure}
\centering
\includegraphics[width=1.00\columnwidth]{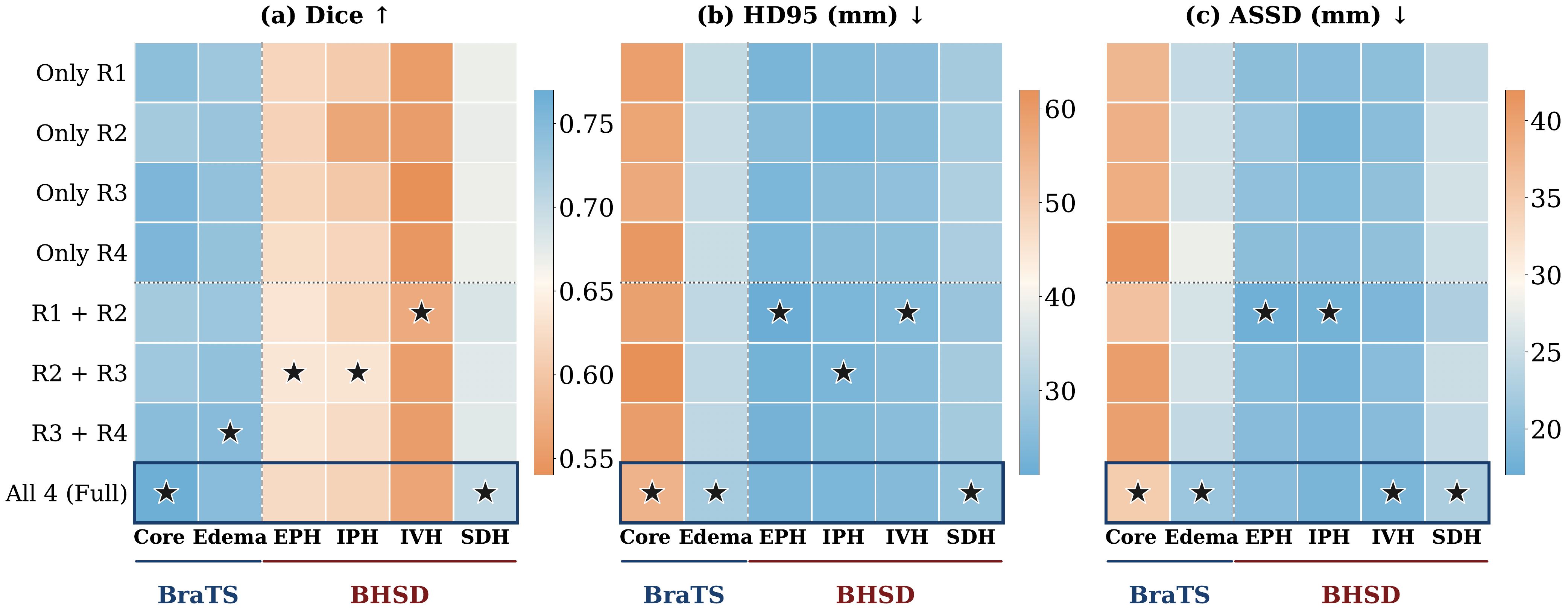}
\caption{Resolution sensitivity analysis on BraTS~2020 and BHSD datasets. (a) Dice, (b) HD95 and (c) ASSD shown across single-scale (R1--R4), two-scale combinations (R1+R2, R2+R3, R3+R4) and full four-scale model (from top to bottom). $\bigstar$ indicates the best result per class; the full four-scale model (black box) achieves the best/near-best performance on both datasets, confirming the contribution of various resolutions to the multi-exit hierarchy.}
\label{fig:resolution_ablation}
\end{figure}

\subsection{Ablation III: Analysis of Multiscale Resolution Sensitivity}
\label{ssec:resolution_results}

Fig.~\ref{fig:resolution_ablation} illustrates segmentation performance across different scale subsets of the four-level multi-exit hierarchy on BraTS 2020 and BHSD datasets. Among single-scale configurations, R3 and R4 achieve equivalent BraTS Tumor Core Dice (0.754), reflecting the semantic richness of deeper features, while R4 attains the highest EPH Dice (0.626), consistent with its capacity to resolve compact hemorrhagic lesions. Two-scale combinations consistently outperform their constituent single scales: R3+R4 achieves Edema Dice of 0.747 and R1+R2 achieves IVH Dice of 0.569, yet no two-scale subset reaches the performance of the full architecture across all classes. The complete four-scale configuration attains the best results on all primary metrics (Tumor Core Dice 0.766, Edema HD95 29.56 mm, SDH Dice 0.704), demonstrating that contributions of multiple scales are complementary rather than redundant, with shallow scales providing spatial precision and deep scales supplying the semantic context, and $\mathcal{A}(\cdot;\theta^c)$ learning the class-specific optimal weighting across scales.

\subsection{Uncertainty Analysis}
\label{ssec:uncertainty_results}
\begin{table}[!tbp]
\centering
\caption{Uncertainty analysis on BraTS~2020 and BHSD datasets via MC Dropout ($T=30$, $p=0.15$) and TTA ($T=12$). $H$: predictive entropy; MI: mutual information; $\sigma_{\mathrm{Dice}}$: Dice uncertainty (standard deviation of Dice across $T$ passes). Subscripts fg/bg denote foreground lesion/background regions respectively.}
\label{tab:uncertainty}
\scriptsize
\setlength{\tabcolsep}{3.6pt}
\renewcommand{\arraystretch}{1.0}
\begin{tabular}{llccccc}
\toprule
& & \multicolumn{5}{c}{\textbf{Epistemic (MC Dropout)}}\\
\cmidrule(r){3-7}
\textbf{Dataset} & \textbf{Class}
& \textbf{$H_{\text{fg}}$} & $H_{\text{bg}}$
& $\text{MI}_{\text{fg}}$ & $\text{MI}_{\text{bg}}$
& $\sigma_{\text{Dice}}$ \\
\midrule
\multirow{3}{*}{BraTS}
& Core
  & 0.140 & 0.168 & 0.0103 & 0.0054 & 0.0056\\
& Edema
  & 0.026 & 0.119 & 0.0065 & 0.0065 & 0.0012 \\
\cmidrule{2-7}
& \textit{Overall}
  & \textit{0.076} & \textit{0.143}
  & \textit{0.0082} & \textit{0.0059} & \textit{0.0034} \\
\midrule
\multirow{5}{*}{BHSD}
& EPH
  & 0.243 & 0.256 & 0.0114 & 0.0081 & 0.0013 \\
& IPH
  & 0.276 & 0.150 & 0.0221 & 0.0062 & 0.0027 \\
& IVH
  & 0.172 & 0.239 & 0.0126 & 0.0083 & 0.0001 \\
& SDH
  & 0.236 & 0.266 & 0.0171 & 0.0080 & 0.0010\\
\cmidrule{2-7}
& \textit{Overall}
  & \textit{0.234} & \textit{0.228}
  & \textit{0.0169} & \textit{0.0076} & \textit{0.0013} \\
\bottomrule
\toprule
& & \multicolumn{5}{c}{\textbf{Aleatoric (TTA)}} \\
\cmidrule(r){3-7} 
\multirow{3}{*}{BraTS}
& Core
  & 0.188 & 0.224 & 0.0119 & 0.0189 & 0.0217 \\
& Edema
  & 0.030 & 0.183 & 0.0085 & 0.0296 & 0.0049 \\
\cmidrule{2-7}
& \textit{Overall}
  & \textit{0.099} & \textit{0.204}
  & \textit{0.0100} & \textit{0.0242} & \textit{0.0133} \\
\midrule
\multirow{5}{*}{BHSD}
& EPH
  & 0.141 & 0.106 & 0.0425 & 0.0431 & 0.0049 \\
& IPH
  & 0.158 & 0.113 & 0.0360 & 0.0260 & 0.0156 \\
& IVH
  & 0.170 & 0.110 & 0.0229 & 0.0412 & 0.0012 \\
& SDH
  & 0.201 & 0.192 & 0.0333 & 0.0297 & 0.0067 \\
\cmidrule{2-7}
& \textit{Overall}
  & \textit{0.154} & \textit{0.130}
  & \textit{0.0326} & \textit{0.0350} & \textit{0.0071} \\
\bottomrule
\end{tabular}
\end{table}

\begin{figure}
\centering
\includegraphics[width=1.00\columnwidth]{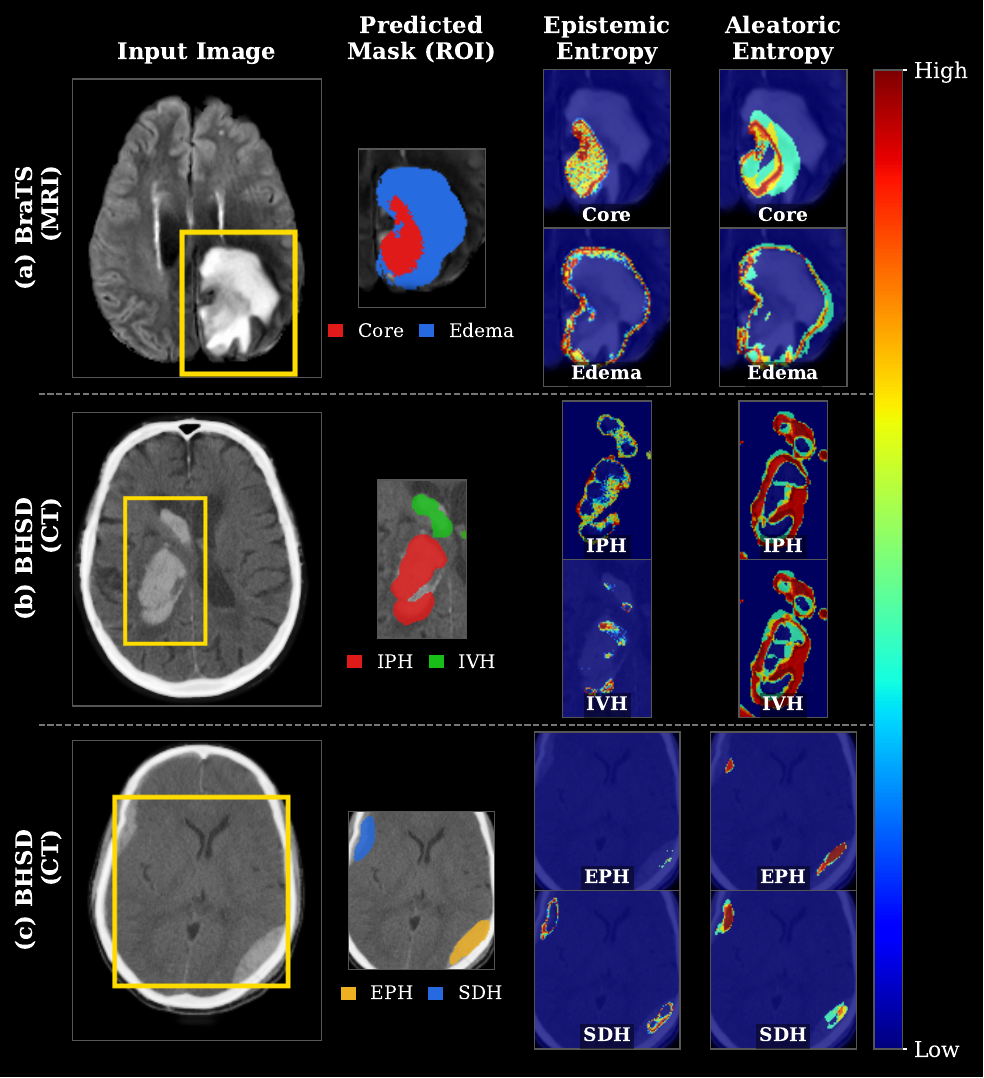}
\caption{Sample uncertainty maps for BraTS~2020 and BHSD datasets. Spatial epistemic (MC Dropout with $T = 30$) and aleatoric (TTA with $T = 12$) entropies for each class prediction on BraTS 2020 (a) and BHSD (b,c), shown within the region of interest (yellow box). Higher uncertainty observed along the lesion boundaries and inter-class competition zones compared to background confirms our results in Table~\ref{tab:uncertainty}.}
\label{fig:uncertainty}
\end{figure}

Table~\ref{tab:uncertainty} reports epistemic and aleatoric uncertainty on BraTS 2020 and BHSD, with visual results in Fig.~\ref{fig:uncertainty}. Under epistemic perturbation, predictive entropy is lower in the foreground than background on BraTS ($H_{\text{fg}} = 0.076$ vs.\ $H_{\text{bg}} = 0.143$), with epistemic MI consistently higher in the foreground on both datasets (BraTS: $0.0082$ vs.\ $0.0059$; BHSD: $0.0169$ vs.\ $0.0076$), confirming that model uncertainty is appropriately concentrated in regions of maximal inter-class competition (also evident from Fig.~\ref{fig:uncertainty}). Among ICH subtypes, IPH exhibits the highest epistemic MI ratio ($\mathrm{MI_{fg}}/\mathrm{MI_{bg}} = 3.56$), reflecting spatial compactness and proximity to competing subtypes, while IVH exhibits the lowest (1.52); Fig.~\ref{fig:uncertainty}(b) illustrates this contrast, with IPH showing diffuse high-entropy activations and IVH uncertainty confined to a narrow boundary ring. Dice uncertainty under epistemic perturbation is negligible ($\sigma_{\text{Dice}} < 0.006$ on BraTS, $< 0.003$ on BHSD), confirming stable segmentation decisions across $T = 30$ stochastic passes. Under aleatoric perturbation, background MI exceeds foreground MI on BraTS ($0.0242$ vs.\ $0.0100$), a direct consequence of $\mathcal{F}_{\text{ag}}^{\text{bin}}$ constraining multiclass predictions within $\Omega^+$; on BHSD, IVH is the most aleatorically stable subtype ($\mathrm{MI_{fg}}/\mathrm{MI_{bg}} = 0.56$) and IPH the least ($1.38$), consistent with the greater spatial ambiguity of parenchymal hemorrhage.

\begin{figure}
\centering
\includegraphics[width=1.00\linewidth]{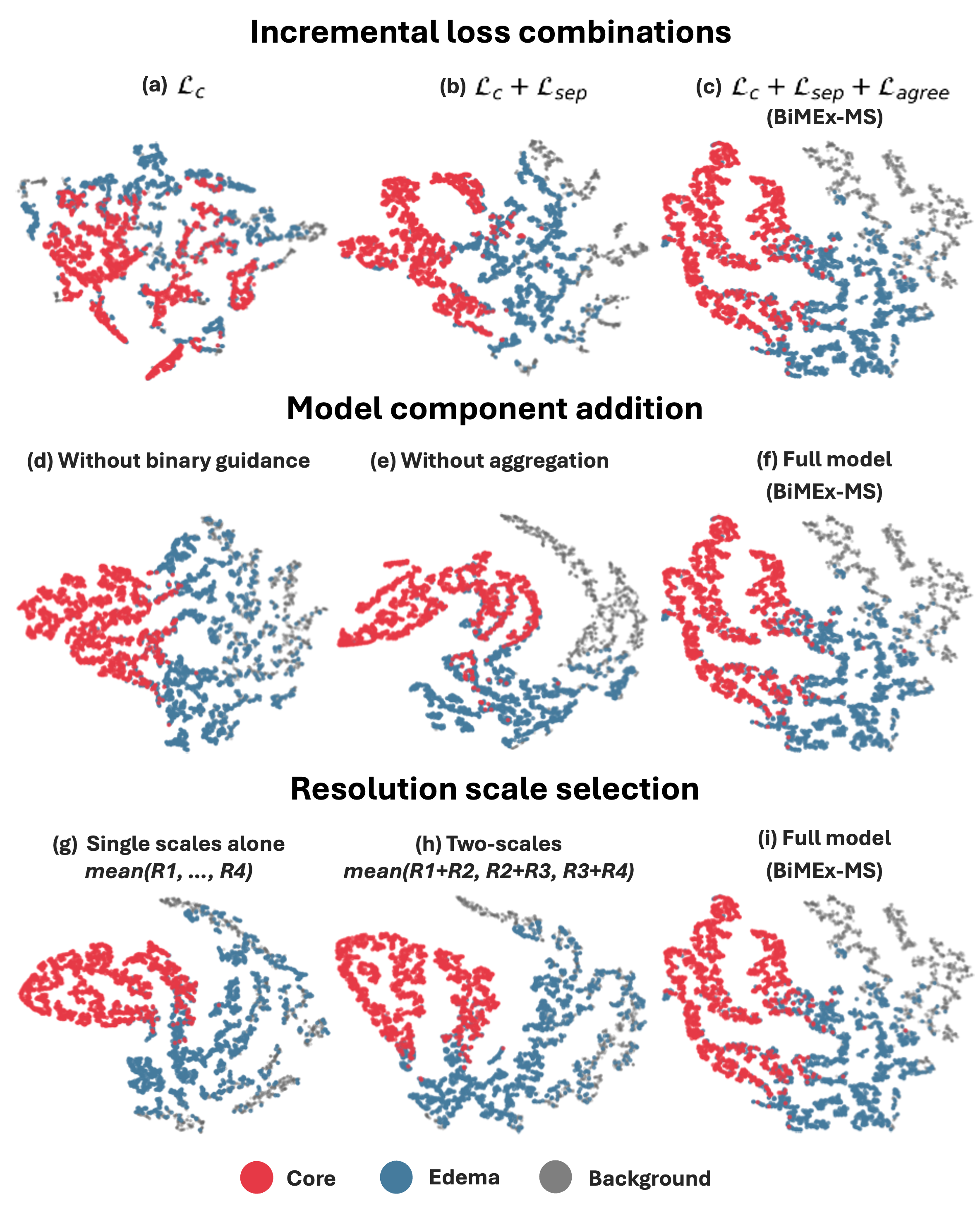}
\caption{t-SNE visualization of learned feature representations 
on the BraTS~2020 dataset. Plots shown for core (red), edema (blue) and background (gray) classes across three ablations: incremental loss combination (row 1), model component ablation (row 2), and resolution scale selection (row 3). Representations of the proposed \textsf{BiMEx-MS} (illustrating better separability with the complete setting) (c,f,i) is shown for reference for each ablation.}
\label{fig:tsne}
\end{figure}

\subsection{Qualitative analysis of representations}
\label{ssec:qualitative_results}

Fig.~\ref{fig:tsne} provides representational evidence complementary to the quantitative ablation results. In row~1, progressive addition of $\mathcal{L}_{\text{sep}}$ and $\mathcal{L}_{\text{agree}}$ successively separates Core and Edema manifolds into geometrically distinct regions, confirming $\mathcal{L}_{\text{sep}}$ as the primary driver of inter-class disentanglement. Row~2 shows that removing binary guidance causes representational collapse, with all three class clusters interleaved and background points dispersed throughout the foreground, reinforcing the role of $\mathcal{F}_{\text{ag}}^{\text{bin}}$ in confining activations to $\Omega^+$. Replacing the aggregation network with uniform averaging partially recovers Core cluster compactness at the expense of Edema fragmentation, consistent with the Edema Dice degradation in Ablation~I. Row~3 demonstrates monotonically improving cluster compactness with increasing scale diversity, with the full four-level hierarchy yielding the sharpest Core--Edema boundary and the greatest background displacement from foreground clusters, confirming that shallow scales contribute spatial resolution and deep scales contribute semantic discriminability. Background displacement from foreground clusters is greatest in the full model, reflecting suppression of false activations outside $\Omega^+$ by $\mathcal{F}_{\text{ag}}^{\text{bin}}$ across all four resolution levels.

\section{Discussion}
\label{sec:discussion}
This study addresses multiclass weakly supervised segmentation of spatially adjacent, co-occurring lesion subtypes in neuroimaging, overcoming background leakage, inter-class blurring, and rare-subtype suppression. \textsf{BiMEx-MS} decouples the problem into a binary localization stream establishing the structural prior $\mathcal{F}_{\text{ag}}^{\text{bin}}$, and a multiclass stream learning class-discriminative representations within $\Omega^+$, consistently outperforming existing weakly supervised baselines with significant improvements in boundary metrics and rare lesion subtypes.

The consistent performance asymmetry between dominant and subordinate classes across all baselines suggests that the core obstacle is not class discriminability but the absence of a mechanism confining inter-class competition to the lesion domain. Consensus-driven, texture-based, and attention-based methods (HAMIL, PathMamba, CoSA) assume intensity contrast absent at tumor core--edema and hemorrhage subtype interfaces, confirming a structural rather than architectural bottleneck. Even MCTformer+'s Contrastive Class Token (Tumor Core Dice 0.754, Table~\ref{tab:brats_results}) induces implicit separation without constraining activation support, producing boundary leakage that $\mathcal{L}_{\text{agree}}$ explicitly addresses. AME-CAM's class-agnostic aggregation causes activation bleed between co-occurring subtypes (e.g., IPH and IVH) (Fig.~\ref{fig:visual_res}), while FMA-WSSS correctly localizes lesion extent but misassigns subtype identity between co-occurring IPH and IVH (Fig.~\ref{fig:visual_res}(d)). This demonstrates that boundary priors alone are insufficient without explicit inter-class exclusivity constraints. The advantage of $\mathcal{F}_{\text{ag}}^{\text{bin}}$ is established by three convergent lines of evidence: its removal causes the largest single performance drop (Fig.~\ref{fig:component_ablation}); t-SNE analysis reveals complete representational collapse in its absence (Fig.~\ref{fig:tsne}, row~2); and improvements are disproportionately large on boundary metrics and rarer classes, consistent with frequency-agnostic spatial support enforced by binary gating.

The incremental ablation (Table~\ref{tab:loss_ablation}, Fig.~\ref{fig:tsne} row~1) reveals that the three loss terms address sequentially dependent failure modes. $\mathcal{L}_c$ establishes per-class foreground--background separation but leaves foreground classes free to collapse onto shared activation directions. $\mathcal{L}_{\text{sep}}$ resolves this via pairwise orthogonality (EPH Dice 0.608, IVH Dice 0.541 on BHSD) but transiently degrades spatial coverage (Tumor Core HD95 60.11 mm), as inter-class exclusion doesn't provide lower bound on activation magnitude within $\Omega^+$. $\mathcal{L}_{\text{agree}}$ closes this gap by grounding the multiclass union in the binary prior, recovering Dice and boundary metrics across all classes. The ordering of loss terms follows from logical dependence, geometrically confirmed by progressive Core--Edema separation in Fig.~\ref{fig:tsne} row~1.

Resolution sensitivity results (Fig.~\ref{fig:resolution_ablation}) confirm that shallow (R1, R2) and deep scales (R3, R4) play complementary roles, however no two-scale subset matches the full hierarchy across all classes. For instance, the shallow scales contribute to the spatial accuracy of the compact subtypes (EPH, IVH), while deep scales (R3, R4) contribute to the semantics of the diffuse subtypes (Edema, SDH). The more consequential finding is the statistically significant Dice drop when $\mathcal{A}(\cdot;\theta^c)$ is replaced by uniform averaging, largest for compact subtypes (EPH, IVH), indicating that rare classes require class-specific scale emphasis that a shared importance map cannot provide.

The RSNA$\,\to\,$BHSD protocol, training on image-level labels and evaluating against independent voxel-level masks without fine-tuning, is the most stringent generalization test. The strong BHSD performance confirms that the binary-guided structural prior transfers across independently acquired CT datasets with different scanner settings and patient demographics. Similarly, the BraTS 2020-trained model evaluated directly on BraTS 2023 SSA, an entirely independent cohort, attains Core Dice 0.759 and Edema Dice 0.711, closely approaching within-dataset results (0.766, 0.745), confirming robust cross-data transfer. The consistent performance across six backbones spanning 3.4M--28.9M parameters (Fig.~\ref{fig:backbone_analysis}), with Swin-Tiny failing to improve upon ResNet-18, confirms that the structural mechanism rather than encoder capacity drives the observed performance improvements.

Progressive expansion results (Table~\ref{tab:class_progression}) provide positive insights in two respects: firstly, introducing EPH as the fourth and rarest class yields bounded, class-local degradation (maximum Dice variation 0.022, HD95 variation below 3 mm) rather than catastrophic collapse. Despite severe underrepresentation, EPH achieves competitive performance in both within-dataset (Dice 0.592) and cross-dataset (Dice 0.562) settings, demonstrating that the class-frequency-agnostic binary prior $\mathcal{F}_{\text{ag}}^{\text{bin}}$ and inter-class orthogonality enforced by $\mathcal{L}_{\text{sep}}$ effectively mitigate long-tailed suppression of rare subtypes. Secondly, the improvement of $+12.4\%$ in the SDH Dice upon introducing IVH, while IPH reduces by $-11.9\%$, indicates that $\mathcal{L}_{\text{sep}}$ redistributes activation contributions across classes rather than degrading all uniformly, yielding a geometrically correct partition of $\Omega^+$. Both IPH and IVH partially recover upon introducing EPH ($+3\%$ each), reflecting continued boundary adjustment as the label space expands.

A notable strength of \textsf{BiMEx-MS} is that its uncertainty estimates are interpretable in terms of structural properties. Epistemic MI is higher in foreground than background on both datasets (BraTS: 0.0082 vs.\ 0.0059; BHSD: 0.0169 vs.\ 0.0076, Table~\ref{tab:uncertainty}), confirming that model uncertainty is anatomically concentrated where inter-class competition is highest (Fig.~\ref{fig:uncertainty}). The aleatoric reversal, where background MI exceeding foreground (BraTS: 0.0242 vs.\ 0.0100), is directly attributable to binary gating suppressing augmentation-induced jitter outside $\Omega^+$. This demonstrates that the structural prior contributes to prediction stability beyond localization accuracy. Also, Dice uncertainty under epistemic perturbation is negligible ($\sigma_{\text{Dice}} < 0.006$ on BraTS), suggesting reliable and anatomically meaningful uncertainty estimates. We consider this a crucial step toward trustworthy deployment under weak supervision. However, prospective radiologist evaluation is required before clinical translation.

\textsf{BiMEx-MS} has a few limitations, that point towards productive future directions. The framework processes 2D axial slices independently, discarding inter-slice context, making the volumetric extension as the most immediate future direction. 
Additionally, pseudo-label quality is inherently bounded by CAM accuracy, a well-understood limitation shared across the WSSS literature. Also, multiclass CT evaluation relies on a single public benchmark (BHSD), reflecting field-wide scarcity of multiclass, multi-label datasets, rather than a framework-specific constraint. Looking ahead, integrating foundation model boundary priors as seeds for $\mathcal{F}_{\text{ag}}^{\text{bin}}$, incorporating vision-language grounding from models such as CLIP to inject pathology-specific semantic constraints, isolating the independent contribution of supervised contrastive pretraining, and prospective clinical validation of uncertainty estimates are the most promising directions for future work.

\section{Conclusion}
In this study, we have addressed mutual exclusivity among co-occurring lesion subregions in multiclass, multi-label weakly supervised neuroimaging segmentation using image-level labels alone.  The proposed binary-guided framework (\textsf{BiMEx-MS}) decouples whole-lesion localization from class-specific discrimination, enforcing exclusivity with per-class separation, inter-class orthogonality, and binary-multiclass spatial consensus. Rigorous validation across two neuroimaging modalities, our method outperforms several weakly supervised baselines. We also perform cross-dataset generalization,  and a progressive four-class expansion, with ablation studies isolating the contribution of each component and uncertainty analysis confirming anatomically meaningful confidence estimates. \textsf{BiMEx-MS} demonstrates that structural guidance rather than model capacity provides the observed improvements. By producing structurally valid, mutually exclusive segmentation masks from image-level labels alone, the framework directly addresses the annotation bottleneck that remains the primary barrier to large-scale clinical deployment of medical image segmentation. Future directions include volumetric extension, vision-language grounding and integration of foundation model priors for improving binary guidance.



\section*{Declaration of competing interest}
The authors declare that they have no known competing financial interests or personal relationships that could have appeared to influence the work reported in this paper.

\section*{Acknowledgments}
This work was supported by DBT/Wellcome Trust India Alliance Fellowship [IA/E/22/1/506763]. This work was also supported in part by a grant from the Council of Scientific \& Industrial Research (CSIR) under its ASPIRE (Women Scientist Scheme) program [25WS(013)/2023-24/EMR-II/ASPIRE] and in part by Start-up Research Grant [SRG/2023/001406] from the Science and Engineering Research Board, India. The authors acknowledge CSR funding from Prazim Trading and Investment Company, which enabled this work. VS is also supported by Pratiksha Trust, Bangalore, India [FG/PTCH-23-1004] and the Seed Research Grant [IE/RERE-22-0583] from the Indian Institute of Science, India.

\bibliographystyle{elsarticle-harv}
\bibliography{references}

\end{document}